# A Deep Learning Density Shaping Model Predictive Gust Load Alleviation Control of a Compliant Wing Subjected to Atmospheric Turbulence

Seid H. Pourtakdoust [1], and Amir H. Khodabakhsh [2]

[1] Corresponding Author, Department of Aerospace Engineering, Sharif University of Technology, Tehran, Iran. Email: pourtak@sharif.edu, ORCID: 0000-0001-5717-6240
[2] Department of Aerospace Engineering, Sharif University of Technology, Tehran, Iran. Email: khodabakhsh@ae.sharif.edu, ORCID: 0000-0002-0457-8673

## Abstract

This study presents a novel deep learning approach aimed at enhancing stochastic Gust Load Alleviation (GLA) specifically for compliant wings. The approach incorporates the concept of smooth wing camber variation, where the camber of the wing's chord is actively adjusted during flight using a control signal to achieve the desired aerodynamic loading. The proposed method employs a deep-learning-based model predictive controller designed for probability density shaping. This controller effectively solves the probability density evolution equation through a custom Physics-Informed Neural Network (PINN) and utilizes Automatic Differentiation for Model Predictive Control (MPC) optimization. Comprehensive numerical simulations were conducted on a compliant wing (CW) model, evaluating performance of the proposed approach against stochastic gust profiles. The evaluation involved stochastic aerodynamic loads generated from Band-Limited White Noise (BLWN) and Dryden gust models. The evaluations were conducted for two distinct Compliant Chord Fractions (CCF). The results demonstrate the effectiveness of the proposed probability density shaping model predictive control in alleviating stochastic gust load and reducing wing tip deflection.

*Keywords*: Deep Learning, Physics-Informed Neural Network, Compliant Wing Gust Load Alleviation, Stochastic Model Predictive Control, Probability Density Shaping

## 1. Introduction

Aviation technology has continuously grown within the past decades. Sustainable aviation, however, demands disruptive ideas for future generations of transport aircraft in terms of safety, fuel and aerodynamic efficiency as well as environmental considerations such as noise and pollution[1]. In addition, there has been a surge in utility of unmanned micro aerial vehicles (UAV) and aerial robotics in recent years. It seems that both aforementioned areas require improvements in a variety of subsystems and disciplines to meet the future challenging issues of safety and environment. Higher aerodynamic and fuel efficiency pushes modern aircraft configurations towards concepts with high aspect ratios and lighter wings that inherently leads designers to more flexible airframes susceptible to large deflections[2]. On the other hand, the complex interaction between the motion of the compliant structure, aircraft control system, inertial and gust loads with aerodynamic forces could potentially lead to flutter, self-excited oscillatory motion, or even wing divergence [3].

Aero-Servo-Elasticity (ASE), initially viewed as undesirable and dangerous in early aviation, can now be utilized to improve flight performance. As an interdisciplinary field, ASE is concerned with dynamic interactions between the control actuation commands, aerodynamic, elastic, and inertial loads of a flexible wing[4]. It has also been noted that at low Reynolds numbers, large portion of the overall drag of micro aerial vehicle is due to flow separation over its flapped or hinged surfaces [5]. While electric actuators commonly used in many aerial vehicles are efficient in terms of weight and power consumption, they occupy significant volume. Seamless aeroelastic wings offer a better alternative actuation for future generations of aerial vehicles. In addition, in many instances electric actuation can be replaced by piezoelectric macro fiber composites that are more efficient in terms of volume, mass, reliability as well as actuation bandwidth [5]. Employing concepts such as the Adaptive Compliant Wing (ACW) for control actuation, active gust load alleviation, flutter, and vibration suppression can also reduce airframes structural weight significantly [6]. Consequently, the development of novel concepts to control the wing structural deflections and/or aeroelastic behavior is of paramount importance [7]. To address this demand, a novel stochastic gust load alleviation controller is proposed in this research.

It is also important to note that despite years of research on aeroelasticity and aero-servo-elastic systems, Uncertainty Quantification (UQ) is addressed only in a handful of studies, most of which focus on parametric uncertainties[8-10]. In this research, however, the UQ of the wing flutter motion in response to stochastic inputs such as the Clear Air Turbulence (CAT) is investigated. The first attempt to analyze the elastic response of an airframe to CAT can be traced back to 1950s[11]. Through decades of research and experiment, different control strategies have been investigated for flutter and vibration suppression induced by stochastic gusts [6, 12, 13]. However, progress in modeling and propagation of uncertainties can lead to safer designs that are also more efficient for active flutter and vibration suppression [3].

A common simplifying assumption in many studies is that the probability density of the wing's response to stochastic loading is Gaussian and the dynamics are linear. This latter assumption leads to stochastic or robust controllers focused on minimizing the mean and/or covariance of desired states [14]. In practice, however, the stochastic input may not necessarily be Gaussian [15]. Furthermore, the nonlinearity associated with the wing response would eventually result in a different evolution for the probability density distributions, even if the input remains Gaussian [16]. Hence, for such systems utility of the mean and variance may not be sufficient to control the stochastic air vehicle response [17]. In this case, potentially better control policies can be synthesized by considering a more comprehensive understanding of the uncertainty specifically the quantified probability density distribution [3].

A Probability density distribution shaping control, or simply Density Shaping Control (DSC), offers a viable solution to address these problems. DSC provides the required flexibility to address a wide range of applications[18].

The primary advantage of DSC lies in its ability to utilize the full probability density information to synthesize control policies, addressing scenarios where traditional mean and variance metrics are insufficient to control stochastic responses. Among many studies on DSC, Ohsumi and Ohtsuka [19] introduced the idea of using the general density evolution equation in a control framework. Following this study, Annunziato and Borzi [20] further expanded the idea and proposed a model predictive DSC framework to control stochastic processes. A key limitation, however, is the inherent requirement to solve the associated governing partial differential equation, the Fokker-Planck-Kolmogorov (FPK) equation, to determine the time evolution of the probability density [21]. This also necessitates assumptions regarding the system's evolution, such as it being a Markovian process, and often presumes gust velocity components can be represented as a generalized white noise process.

Recently, FPK-DP Net , a novel physics informed deep learning approach, is proposed to solve the time evolution of the probability density of stochastic systems via a Physics-Informed Neural Network (PINN) [22]. Employing the DeepPDEM [23] concept, FPK-DP Net solves a dimension-reduced form of the Fokker-Planck-Kolmogorov (FPK) equation. FPK-DP Net offers several advantages for solving the FPK equation. As a Physics-Informed Machine Learning (PIML) framework, it uniquely solves the time evolution of probability density without requiring large quantities of labeled training data. Instead, FPK-DP Net learns the FPK equation by minimizing the differential equation in its residual form using randomly selected collocation points from the domain. Furthermore, FPK-DP Net's foundation in Deep Neural Networks (DNNs) means it leverages their capability as universal function and operator approximators for complex nonlinear mappings, and utilizes Automatic Differentiation (AD) for accurate derivative calculations. Utilizing AD helps avoiding the computational expense and potential errors of traditional numerical differentiation. However, like other deep learning methods, its gradient-based optimizers may converge to local optima, and determining the optimal network architecture, such as the number of neurons, can be a challenging, problem-dependent task.

Utilizing the FPK-DP Net in this research, a novel density shaping controller is proposed for vibration suppression of a CW. To this aim, the governing equations of motion of a compliant aeroelastic wing are first summarized. Subsequently, a detailed description of the proposed Density Shaping Model Predictive Controller (DSMPC) is provided. Finally, the proposed method is numerically implemented on a typical compliant wing and its performance is investigated. The core novelty of this research is the integration of the FPK-DP Net, a novel PINN framework, to efficiently solve a dimension-reduced form of the FPK equation for the time-evolution of probability density within an MPC scheme for aeroelastic applications. The main objectives of this research are summarized as follows:

- To propose a novel stochastic gust load alleviation controller for future generations of transport aircraft
- To investigate the UQ of wing flutter motion in response to stochastic inputs, such as CAT
- To propose and numerically implement a novel DSMPC for the vibration suppression of a compliant wing, utilizing the FPK-DP Net for calculating probability density distribution evolution
- To assess the performance of the proposed method for controlling wing structural deflections and aeroelastic behavior under stochastic excitations

The proposed framework holds promise for real-world applications. By enabling control over the probability density distribution of the states of the wing, this research can contribute to safer and more efficient designs for next-generation aircraft, including morphing and high-aspect-ratio wings.

## 2. Aero-Servo-Elastic Model

This research investigates a smooth wing camber variation concept utilizing a novel vibration suppression control methodology. To this aim, an unswept, untapered cantilever wing with a relatively high aspect ratio is assumed. In addition, the mass distribution of the wing is presumed to be uniform. Therefore, the one-dimensional beam approximation is considered as an adequate structural model [24] for our conceptual investigation. In addition, it is reasonably assumed that below transonic speeds the flow remains mostly attached and the AoA remains relatively small in cruising flights [2]. Furthermore, the unperturbed freestream velocity is also assumed to be much larger than any component of turbulence. Considering "Taylor's frozen turbulence hypothesis", the turbulence velocities can be presumed constant as they pass over the wing [25]. Given the aforementioned assumptions, the classical aeroelastic model and the Lagrangian mechanics can be used to model the aeroservoelastic behavior of the CW [4].

## 2.1 Structural Dynamics of the Compliant Wing

The wing is modeled as a cantilever uniform Euler-Bernoulli beam. This model, utilized as a one-dimensional approximation for this conceptual investigation, inherently assumes a slender geometry, small deflections, and neglects shear deformation and rotary inertia. These assumptions are considered appropriate given the context and the assumed geometry. Let $EI$ and $GJ$ denote bending and torsional rigidity of the CW, respectively. The strain energy "$U$" can be written as

$$U = \frac{1}{2}\int_0^\ell (EI(w'')^2 + GJ(\theta')^2)dy \tag{1}$$

where $\ell$, $w$ and $\theta$ are the wing span, vertical, and twisting deflection of the wing, respectively. Denoting mass per unit volume of the material by "$\rho$", the kinetic energy "$K$" of the CW model would be,

$$K = \frac{1}{2}\int_0^\ell \iint_A \rho\left((\dot{w} - x\dot{\theta})^2 + z^2\dot{\theta}^2\right) dx\, dz\, dy \tag{2}$$

After evaluation of the integrals Eq. (2) yields,

$$K = \frac{1}{2}\int_0^\ell m \begin{Bmatrix}\dot{w}\\ \dot{\theta}\end{Bmatrix}^T \begin{bmatrix} 1 & -bx_\theta \\ -bx_\theta & b^2r^2\end{bmatrix} \begin{Bmatrix}\dot{w}\\ \dot{\theta}\end{Bmatrix} dy \tag{3}$$

Let $\bar{M} = M_{\frac{1}{4}} + \left(a + \frac{1}{2}\right)bL$, the virtual work would then be ,

$$\delta W = \int_0^\ell L\delta w + \bar{M}\delta\theta\, dy \tag{4}$$

From the Lagrangian mechanics, the EoM can be deduced as,

$$\frac{d}{dt}\left(\frac{\partial K}{\partial \dot{\xi}_i}\right) + \frac{\partial U}{\partial \xi_i} = \Xi_i \tag{5}$$

$$\int_0^\ell m \begin{bmatrix} 1 & -bx_\theta \\ -bx_\theta & b^2 r^2 \end{bmatrix} \begin{Bmatrix} \ddot{w} \\ \ddot{\theta} \end{Bmatrix} dy + \int_0^\ell \begin{bmatrix} EI & 0 \\ 0 & 0 \end{bmatrix} \begin{Bmatrix} w'' \\ \theta'' \end{Bmatrix} + \begin{bmatrix} 0 & 0 \\ 0 & GJ \end{bmatrix} \begin{Bmatrix} w' \\ \theta' \end{Bmatrix} dy = \begin{Bmatrix} \Xi_w \\ \Xi_\theta \end{Bmatrix} \tag{6}$$

where $b$, $x_\theta$, $r$, and $\Xi$ denote semi-chord, static-unbalance parameter, the dimensionless radius of gyration of the section, and the generalized force, respectively. Utilizing the assumed modes method and assuming free-vibration modes for the cantilever wing leads to a solution,

$$w(t,y) = \{\mathbf{\Psi}(y)\}^T\{\mathbf{\eta}(t)\}, \quad \theta(t,y) = \{\mathbf{\Theta}(y)\}^T\{\mathbf{\phi}(t)\} \tag{7}$$

where,

$$\Psi_j(y) = \cosh(\alpha_j y) - \cos(\alpha_j y) - \beta_j(\sinh(\alpha_j y) - \sin(\alpha_j y)) \tag{8}$$

$$\Theta_i(y) = \sqrt{2} \sin\left(\frac{\pi\left(i - \frac{1}{2}\right)}{\ell} y\right) \tag{9}$$

Let $A_{ij} = \frac{1}{\ell}\int_0^\ell \Theta_i \Psi_j dy$. Substituting the assumed modes into the virtual work expression, the strain, and the kinetic energy relations, the Lagrange's equation yields,

$$m\ell \begin{bmatrix} [\mathbf{I}] & -bx_\theta[\mathbf{A}]^T \\ -bx_\theta[\mathbf{A}] & b^2 r^2[\mathbf{I}] \end{bmatrix} \begin{Bmatrix} \ddot{\mathbf{\eta}} \\ \ddot{\mathbf{\phi}} \end{Bmatrix} + \ell \begin{bmatrix} EI\,\text{diag}(\mathbf{\alpha})^4 & 0 \\ 0 & GJ\,\text{diag}(\mathbf{\gamma})^2 \end{bmatrix} \begin{Bmatrix} \mathbf{\eta} \\ \mathbf{\phi} \end{Bmatrix} = \begin{Bmatrix} \Xi_w \\ \Xi_\theta \end{Bmatrix} \tag{10}$$

Let $L$ and $\bar{M}$ denote the aerodynamic lift and moment, respectively. The generalized forces are,

$$\{\Xi_w\} = \int_0^\ell \{\mathbf{\Psi}\} L \, dy, \quad \{\Xi_\theta\} = \int_0^\ell \{\mathbf{\Theta}\} \bar{M} \, dy \tag{11}$$

## 2.2 Aerodynamic Modeling

In accordance with Peter's method [26], a finite state representations is adopted for the aerodynamic forces and moments. Let $w_G$ and $q_G$ represent linear vertical and angular velocities induced by the stochastic gust field. Following Peters' method [26], the expressions for the aerodynamic forces and moments can be represented as,

$$L = \pi\rho_\infty b\left(-b[1 \quad ab]\left(\begin{Bmatrix}\ddot{w}\\\ddot{\theta}\end{Bmatrix}+\begin{Bmatrix}\dot{w}_G\\\dot{q}_G\end{Bmatrix}\right)+2U[-1 \quad b(1-a)]\left(\begin{Bmatrix}\dot{w}\\\dot{\theta}\end{Bmatrix}+\begin{Bmatrix}w_G\\q_G\end{Bmatrix}\right)+2U^2[0 \quad 1]\begin{Bmatrix}w\\\theta\end{Bmatrix}\right.$$
$$\left.-2U\lambda_0\right) \tag{12}$$

$$\bar{M} = \pi\rho_\infty b^2\left(-b\left[a \quad b\left(a^2+\frac{1}{8}\right)\right]\left(\begin{Bmatrix}\ddot{w}\\\ddot{\theta}\end{Bmatrix}+\begin{Bmatrix}\dot{w}_G\\\dot{q}_G\end{Bmatrix}\right)-U[2a+1 \quad ab(2a-1)]\left(\begin{Bmatrix}\dot{w}\\\dot{\theta}\end{Bmatrix}+\begin{Bmatrix}w_G\\q_G\end{Bmatrix}\right)\right.$$
$$\left.+(2a+1)U^2[0 \quad 1]\begin{Bmatrix}w\\\theta\end{Bmatrix}-(2a+1)U\lambda_0\right) \tag{13}$$

where $\rho_\infty$ and $U$ are the free-stream density and velocity, respectively. The average induced-flow velocity $\lambda_0$ can be approximated in terms of $N_\lambda$ induced flow states $\lambda_i$ as,

$$\lambda_0 \approx \frac{1}{2}\sum_{i=1}^{N_\lambda}b_i\lambda_i = \frac{1}{2}\{b\}^T\{\lambda\}, \qquad b_i = \begin{cases}(-1)^{i-1}\dfrac{(N_\lambda+i-1)!}{(N_\lambda-i-1)!\,(i!)^2} & i \neq N_\lambda \\ (-1)^{i-1} & i = N_\lambda\end{cases} \tag{14}$$

By defining $c_i = \frac{i}{2}$ and choosing an appropriate $N_\lambda$, the induced-flow states can be approximated by solving an ODE system,

$$[\mathcal{A}]\{\dot{\lambda}\} = -\frac{U}{b}\{\lambda\} + \{c\}\left([-1 \quad b\left(\frac{1}{2}-a\right)]\begin{Bmatrix}\ddot{w}\\\ddot{\theta}\end{Bmatrix}+[0 \quad U]\begin{Bmatrix}\dot{w}\\\dot{\theta}\end{Bmatrix}\right) \tag{15}$$

Now let,

$$[\overline{\mathcal{A}_1}] = -\frac{U}{b}[\mathcal{A}]^{-1} \tag{16}$$

Since Peter's method uses the averaged induced velocity,

$$[\overline{\mathcal{A}_2}] = \frac{1}{\ell}[\mathcal{A}]^{-1}\{c\}[0 \quad U]\int_0^\ell\begin{bmatrix}\{\Psi\}^T & 0\\ 0 & \{\Theta\}^T\end{bmatrix}dy \tag{17}$$

And,

$$[\overline{\mathcal{A}_3}] = \frac{1}{\ell}[\mathcal{A}]^{-1}\{c\}\left[-1 \quad b\left(\frac{1}{2}-a\right)\right]\int_0^\ell\begin{bmatrix}\{\Psi\}^T & 0\\ 0 & \{\Theta\}^T\end{bmatrix}dy \tag{18}$$

Which leads to,

$$\{\dot{\lambda}\} = [\mathcal{A}_1]\{\lambda\} + [\mathcal{A}_2]\begin{Bmatrix}\dot{\eta}\\\dot{\phi}\end{Bmatrix} + [\mathcal{A}_3]\begin{Bmatrix}\ddot{\eta}\\\ddot{\phi}\end{Bmatrix} \tag{19}$$

Other matrices and vectors in Eq. (15) are defined as,

$$[\mathcal{A}] = [\mathbf{D}] + \{d\}\{b\}^T + \{c\}\{d\}^T + \frac{1}{2}\{c\}\{b\}^T \tag{20}$$

$$D_{ij} = \begin{cases} \dfrac{1}{2i} & i = j+1 \\ -\dfrac{1}{2i} & i = j-1 \\ 0 & i \neq j \pm 1 \end{cases} \tag{21}$$

$$d_i = \begin{cases} \dfrac{1}{2} & i = 1 \\ 0 & i \neq 1 \end{cases} \tag{22}$$

$$c_i = \frac{i}{2} \tag{23}$$

Hence, the generalized forces can be rewritten as,

$$L = \pi\rho_\infty b \left(-b[1 \quad ab]\begin{Bmatrix}\ddot{w}\\\ddot{\theta}\end{Bmatrix} + 2U[-1 \quad b(1-a)]\begin{Bmatrix}\dot{w}\\\dot{\theta}\end{Bmatrix} + 2U^2[0 \quad 1]\begin{Bmatrix}w\\\theta\end{Bmatrix} - 2U\lambda_0\right) \tag{24}$$

$$\begin{aligned}\bar{M} = \pi\rho_\infty b^2 &\left(-b\left[a \quad b\left(a^2 + \frac{1}{8}\right)\right]\begin{Bmatrix}\ddot{w}\\\ddot{\theta}\end{Bmatrix} - U[2a+1 \quad ab(2a-1)]\begin{Bmatrix}\dot{w}\\\dot{\theta}\end{Bmatrix}\right.\\ &\left.+ (2a+1)U^2[0 \quad 1]\begin{Bmatrix}w\\\theta\end{Bmatrix} - (2a+1)U\lambda_0\right)\end{aligned} \tag{25}$$

## 2.3 Compliant Chord Representation

Assuming a fishbone configuration [27] as depicted in Figure 1, the convention used by Zhang et al. [28] s adopted to model the smooth camber variation concept. To this end, the section is assumed to consist of two parts: rigid and compliant sections. The compliant section of the airfoil is modeled as a cantilevered uniform Euler-Bernoulli beam.

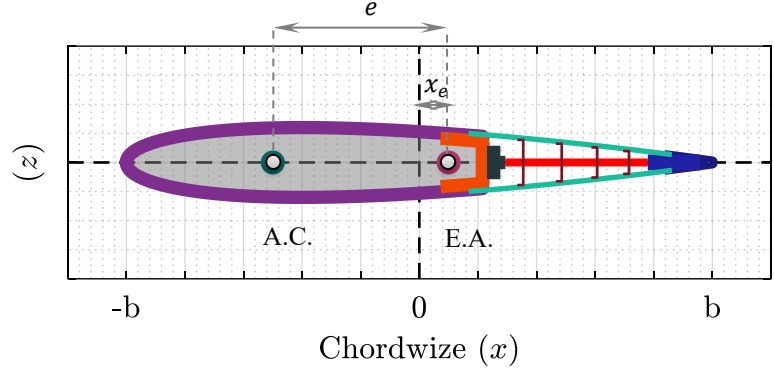

Figure 1 - Schematic representation of the compliant wing section with smooth camber variation

Let subscript "$c$" denote the compliant chord parameters. Considering only the deformation caused by the actuator, and neglecting other factors, the potential energy of the compliant chord can be represented as,

$$\int_0^{\ell_c} \left(EI_c w_c^{(4)} + \mu_c \ddot{w}_c\right) dx = \int_0^{\ell_c} q_c \, dx \tag{26}$$

where $q_c$ represents the control moment applied to the compliant section. Let $q_c = \bar{u}_1 \delta(x - x_1) + \bar{u}_2 \delta(x - x_2)$, hence

$$\{\Xi_c\} = \bar{u}_1 \{\Psi_c(x_1)\} + \bar{u}_2 \{\Psi_c(x_2)\}, \qquad \Xi_{c_j} = \bar{u}_1 \Psi_j^c(x_1) + \bar{u}_2 \Psi_j^c(x_2) \tag{27}$$

Therefore,

$$\{\ddot{\eta}_c\} = -\frac{EI_c}{m_c} \mathrm{diag}(\boldsymbol{\alpha}_c)^4 \{\boldsymbol{\eta}_c\} + \frac{1}{m_c \ell_c} \mathrm{diag}([\{\Psi_c(x_1)\}, \{\Psi_c(x_2)\}]) \begin{Bmatrix} \bar{u}_1 \\ \bar{u}_2 \end{Bmatrix} \tag{28}$$

The aerodynamic force and moment due to the camber variation can be found by applying the thin airfoil theory and integrating the pressure difference over the compliant chord. Assuming small chord curvatures, i.e., $\cos\left(\frac{\partial w_c}{\partial x}\right) \approx 1 - \frac{1}{2}\left(\frac{\partial w_c}{\partial x}\right)^2 \approx 1$, the pressure difference is,

$$\Delta P = -2\rho_\infty \ddot{w}_c \sqrt{x(\ell_c - x)} + \frac{\rho_\infty U(2x - \ell_c)}{\sqrt{x(\ell_c - x)}} (\dot{w}_c + U w_c') \tag{29}$$

$$L_c = \int_0^{\ell_c} \cos(w_c') \Delta P \, dx, \qquad M_c = \int_0^{\ell_c} x \cos(w_c') \Delta P \, dx \tag{30}$$

## 3. Density Shaping Control

This section introduces the framework of Density Shaping Control (DSC), explaining its theoretical foundations and how it is applied to compliant wing gust load alleviation. Presuming that the gust velocity components can be represented as a generalized white noise process $\mathbf{W}_t$ [29, 30], the equations of motion for the CW can be reformulated as an Itô process $\{\mathbf{X}_t\}_{t\in T} \in \mathbb{R}^n$ over some probability space $(\Omega, \mathcal{F}, P)$ such that [16],

$$\dot{\mathbf{X}}_t = \mathbf{f}(t, \mathbf{X}_t; u) + \mathbf{g}(t, \mathbf{X}_t; u)\mathbf{W}_t(\omega) \quad t \geq 0 \tag{31}$$

where $\mathbf{X}$ is a uniformly bounded random process. $\mathbf{f}: \mathbb{R}^+ \times \mathbb{R}^n \to \mathbb{R}^n$ denotes the deterministic behavior of the wing also known as the drift vector. $\mathbf{g}: \mathbb{R}^+ \times \mathbb{R}^n \to \mathbb{R}^n \times \mathbb{R}^m$ is the gust influence matrix also identified as the diffusion or dispersion matrix. $\mathbf{W}_t(\omega)$ represents an m-dimensional of the gust modelled as white noise, and $u$ is the control signal. Let $p_\mathbf{X}(t, \mathbf{x}): \mathbb{R}^+ \times \mathbb{R}^n \to \mathbb{R}^+$ denote the instantaneous joint probability density distribution of the system satisfying the Kolmogorov axioms [31],

$$\int_{\Omega_\mathbf{x}} p_\mathbf{X}(t, \mathbf{x}) d\mathbf{x} = 1 \quad \wedge \quad p_\mathbf{X}(t, \mathbf{x}) > 0, \forall \mathbf{x} \in \mathbb{R}^n \tag{32}$$

Without loss of generality, assume that the Initial Condition (IC), i.e., the initial joint probability density of the states $p_\mathbf{X}(0, \mathbf{x})$, is known and independent of the stochastic input process. The time evolution of the joint PDF $p_\mathbf{X}(t, \mathbf{x})$ is governed by the Fokker-Planck-Kolmogorov (FPK) equation [22, 31].

$$\partial_t p_\mathbf{X}(t, \mathbf{x}) = -\nabla_\mathbf{x} \cdot \left(\mathbf{f}(t, \mathbf{x}) p_\mathbf{X}(t, \mathbf{x})\right) + \mathbf{D}(t, \mathbf{x}) : \mathcal{H}_\mathbf{x}\left(p_\mathbf{X}(t, \mathbf{x})\right) \tag{33}$$

Where ":" is the Frobenius inner product (double-contraction product), $\mathcal{H}_\mathbf{x}(\cdot)$ is the Hessian matrix and $\mathbf{D}(\mathbf{x}, t)$ is the diffusion tensor defined as $\frac{1}{2}\mathbf{g}\mathbf{g}^T$. To obtain a well-posed problem in an unbounded domain, a homogeneous Boundary Condition (BC) at infinity (far-field vanishing condition) is also assumed [32].

$$\lim_{\|\mathbf{x}\|\to\infty} f_i(t, \mathbf{x}) p_\mathbf{X}(t, \mathbf{x}) = \lim_{\|\mathbf{x}\|\to\infty} D_{ij} p_\mathbf{X}(t, \mathbf{x}) = \lim_{\|\mathbf{x}\|\to\infty} \partial_{x_i} D_{ij} p_\mathbf{X}(t, \mathbf{x}) = 0 \tag{34}$$

The system presented in Eq. (33) is localized. Therefore, both the IC and its time evolution are of compact support. One should note that the probability density transport modeled as the FPK equation implicitly assumes that $\mathbf{X}_t$ evolves continuously or at least by small jumps, and can be approximated by a Markovian process [33]. Suppose that the control input $u$ is known, finite, and only updated at discrete instances of time $\tau_k$. Thus, Eq. (33) can be rewritten such that it represents the evolution of PDF starting at some instance of discrete-time $\tau_k$.

$$\partial_t p_{\mathbf{X}|u_k}(t, \mathbf{x}; u_k) = -\nabla_{\mathbf{x}} \cdot \left(\mathbf{f}(t, \mathbf{x}; u_k) p_{\mathbf{X}|u_k}(t, \mathbf{x}; u_k)\right) + \mathbf{D}(t, \mathbf{x}; u_k) : \mathcal{H}_{\mathbf{x}}\left(p_{\mathbf{X}|u_k}(t, \mathbf{x}; u_k)\right), \quad (35)$$
$$t \in (\tau_k, \tau_k + \hbar]$$

Let $\mathcal{D}$ represent the FPK differential operator such that:

$$\mathcal{D}_k p_{\mathbf{X}|u_k}(t, \mathbf{x}; u_k) = 0, \qquad t \in (\tau_k, \tau_k + \hbar] \quad (36)$$

where the subscript $k$ denotes the realization of the functions when the control input $u_k$ is constant. Eq. (36) also has a homogeneous BC at infinity and its IC is determined recursively presuming $p_{\mathbf{X}|u_{-1}}(\tau_0, \mathbf{x}) = p_{\mathbf{X}}(0, \mathbf{x}),$.

$$p_{\mathbf{X}|u_k}(\tau_k, \mathbf{x}) = p_{\mathbf{X}|u_{k-1}}(\tau_k, \mathbf{x}) \ \wedge \ p_{\mathbf{X}|u_0}(0, \mathbf{x}) = p_{\mathbf{X_0}}(\mathbf{x}) \quad (37)$$

## 3.1 Model Predictive Density Shaping Scheme

In this study the Receding Horizon Model Predictive Control (RHMPC) concept [34] is employed to develop a probability density shaping control. The objective function is commonly defined as a metric (distance function) between the desired and the realized states. The Kullback-Leibler divergence or the cross-entropy [35] and the standard $L^2$ norm [20, 36] are the two metrics suggested in the literature to quantify the distance between probability distributions. Assuming one time-step as the control horizon, let $p_{\mathbf{X}}^d(T, \mathbf{x})$ denote the desired joint probability density of the states over the prediction horizon $T = [\tau_k, \tau_k + \hbar)$. Assuming that the probability density distributions are and remain Lebesgue integrable, the $\mathcal{L}^2$ norm is used as the metric between the desired PDF $p_{\mathbf{X}}^d(T, \mathbf{x})$ and the predicted one $p_{\mathbf{X}}(T, \mathbf{x}; u_k)$ over the time interval $T$. Thus, the objective function can be defined similar to the classic Bolza problem. The objective function includes only the Lagrangian term, neglecting the Mayer (terminal) term,

$$\mathcal{J}(u_k) = \int_{\tau_k}^{\tau_k+\hbar} \int_{\Omega_{\mathbf{x}}} \left( \frac{1}{2} \left( \mathcal{P}_{\mathbf{X}|u_k}(t,\mathbf{x};u_k) - \mathcal{P}_{\mathbf{X}}^d(t,\mathbf{x}) \right)^2 + \frac{c}{2} u_k^2 \right) d\mathbf{x}\, dt \qquad (38)$$

where "$c$" is a tunable parameter. The optimal control problem can now be formulated as:

$$u_k^* = \arg\min_{u_k} \mathcal{J} \qquad (39)$$
$$s.t.: \mathcal{D}_k \mathcal{P}_{\mathbf{X}|u_k}(t,\mathbf{x};u_k) = 0$$

## 3.2 Optimization Strategy

It is reasonable to assume that both the probability density distribution and the $L^2$-norm metric are continuous, Lipschitz and at least twice differentiable with respect to the arguments. These assumptions imply that Eq. (39) admits a unique solution: an optimal constant control action $u_k$ for each time interval. This action is calculated by optimizing the model over the prediction horizon $\hbar$ while satisfying the FPK equation as an equality constraint. Suppose that the Leibniz integral rule in the sense of the measure theory holds for $\mathcal{P}_{\mathbf{X}}$ [22]. Let "$l$" denote the Lagrangian multiplier. Therefore, the auxiliary Lagrangian function $\mathcal{L}$ is,

$$\mathcal{L} = \frac{1}{2} \left( \mathcal{P}_{\mathbf{X}|u_k}(t,\mathbf{x};u_k) - \mathcal{P}_{\mathbf{X}}^d(t,\mathbf{x}) \right)^2 + \frac{c}{2} u_k^2 + l \cdot \mathcal{D}_k \mathcal{P}_{\mathbf{X}|u_k}(t,\mathbf{x};u_k) \qquad (40)$$

Hence, the Karush–Kuhn–Tucker (KKT) necessary conditions are,

$$\begin{cases} \dfrac{\partial \mathcal{L}}{\partial u_k} = 0 \\ l \cdot \left( \mathcal{D}_k \mathcal{P}_{\mathbf{X}|u_k}(t,\mathbf{x};u_k) \right) = 0 \\ \mathcal{D}_k \mathcal{P}_{\mathbf{X}|u_k}(t,\mathbf{x}) = 0 \\ l \geq 0 \end{cases} \qquad (41)$$

A possible solution to find the optimal control input is to use gradient-based optimizers. The first-order gradient-based optimizers, such as the steepest descent only use the first derivative of the objective function with respect to the optimization variable while the second-order methods utilize the second-order derivatives, as well[37]. Utilizing the interior-point algorithm[38], the optimization problem can be solved efficiently assuming that both the first and the second derivatives of the Lagrangian with respect to the control input are available. In the following expressions, the arguments of the functions are omitted for the sake of brevity.

$$\frac{\partial \mathcal{L}}{\partial u_k} = \left(p_{\mathbf{X}|u_k} - p_{\mathbf{X}}^d\right)\frac{\partial p_{\mathbf{X}|u_k}}{\partial u_k} + 2cu_k + \lambda \cdot \frac{\partial}{\partial u_k}\left(\mathcal{D}_k p_{\mathbf{X}|u_k}\right) \tag{42}$$

$$\frac{\partial^2 \mathcal{L}}{\partial u_k^2} = \left(p_{\mathbf{X}|u_k} - p_{\mathbf{X}}^d\right)\frac{\partial^2 p_{\mathbf{X}|u_k}}{\partial u_k^2} + \left(\frac{\partial p_{\mathbf{X}|u_k}}{\partial u_k}\right)^2 + 2c + \lambda \cdot \frac{\partial^2}{\partial u_k^2}\left(\mathcal{D}_k p_{\mathbf{X}|u_k}\right) \tag{43}$$

The derivatives of the FPK equation can also be represented as,

$$\begin{aligned}&\frac{\partial}{\partial u_k}\mathcal{D}_k p_{\mathbf{X}|u_k} \\ &= \frac{\partial^2 p_{\mathbf{X}|u_k}}{\partial t \partial u_k} + \nabla_{\mathbf{x}} \cdot \left(\frac{\partial \mathbf{f}}{\partial u_k}p_{\mathbf{X}|u_k} + \mathbf{f}\frac{\partial p_{\mathbf{X}|u_k}}{\partial u_k}\right) - \frac{\partial \mathbf{D}}{\partial u_k}:\mathcal{H}_{\mathbf{x}}\left(p_{\mathbf{X}|u_k}\right) - \mathbf{D}:\mathcal{H}_{\mathbf{x}}\left(\frac{\partial p_{\mathbf{X}|u_k}}{\partial u_k}\right)\end{aligned} \tag{44}$$

$$\begin{aligned}&\frac{\partial^2}{\partial u_k^2}\mathcal{D}_k p_{\mathbf{X}|u_k} \\ &= \frac{\partial^3 p_{\mathbf{X}|u_k}}{\partial t \partial u_k^2} + \nabla_{\mathbf{x}} \cdot \left(\frac{\partial^2 \mathbf{f}}{\partial u_k^2}p_{\mathbf{X}|u_k} + 2\frac{\partial \mathbf{f}}{\partial u_k}\frac{\partial p_{\mathbf{X}|u_k}}{\partial u_k} + \mathbf{f}\frac{\partial^2 p_{\mathbf{X}|u_k}}{\partial u_k^2}\right) - \frac{\partial^2 \mathbf{D}}{\partial u_k^2}:\mathcal{H}_{\mathbf{x}}\left(p_{\mathbf{X}|u_k}\right) \\ &\quad - 2\frac{\partial \mathbf{D}}{\partial u_k}:\mathcal{H}_{\mathbf{x}}\left(\frac{\partial p_{\mathbf{X}|u_k}}{\partial u_k}\right) - \mathbf{D}:\mathcal{H}_{\mathbf{x}}\left(\frac{\partial^2 p_{\mathbf{X}|u_k}}{\partial u_k^2}\right)\end{aligned} \tag{45}$$

These derivatives can be used to solve the KKT equations via Newton's steps [38] and Cholesky's decomposition.

$$\begin{bmatrix} \frac{\partial^2 \mathcal{L}}{\partial u_k^2} & \frac{\partial \mathcal{D}_k p_{\mathbf{X}|u_k}}{\partial u_k} \\ \frac{\partial \mathcal{D}_k p_{\mathbf{X}|u_k}}{\partial u_k} & 0 \end{bmatrix}\begin{bmatrix} \Delta u_k \\ \Delta \lambda \end{bmatrix} = -\begin{bmatrix} \frac{\partial \mathcal{L}}{\partial u_k} \\ \mathcal{D}_k p_{\mathbf{X}|u_k} \end{bmatrix} \tag{46}$$

Even though the second-order methods usually converge faster, the numerical determination of the second-order derivatives is computationally very expensive[39]. As a novel concept, however, we use a neural network model as the predictive model and utilize Automatic Differentiation. This technique, used in neural networks almost since their inception, allows finding derivatives using the chain rule[23]. Algorithm 1 provides a pseudo-code overview of the proposed DSMPC for compliant wing gust load alleviation.

Algorithm 1 DSMPC with FPK-DP Net for Compliant Wing Gust Load Alleviation

**Require:** Wing parameters (Table 1), $n_s = 10$ structural modes, $n_a = 6$ aerodynamic states, gust model (BLWN/Dryden), $3\sigma_w = 5 \ m/s$, $\alpha_{\text{root}} = 4°$, $\Delta t = 0.01$, $|M| \leq 50 \ N.m$

**Ensure:** States $\mathbf{x}(t)$, control $u(t)$, probability distribution densities $p_\mathbf{X}(t, \mathbf{x})$.

1. Assemble modal EoM (Eqs. (6)-(11)), aero model (Eqs. (12)-(25)), compliant chord (Eqs. (26)-(30)).
2. Load pretrained FPK-DP Net (Eqs. (48)-(56))
3. Initialize $\mathbf{x}(0)$ at trim, $t \leftarrow 0$, $u_{-1} \leftarrow 0$.
4. **while** $t < T_{\text{sim}}$ **do**
5.     Calculate current state $\mathbf{x}$
6.     Predict probability density via FPK-DP Net

$$\partial_t p_{\mathbf{X}|u_k}(t, \mathbf{x}; u_k) = -\nabla_\mathbf{x} \cdot \left(\mathbf{f}(t, \mathbf{x}; u_k) p_{\mathbf{X}|u_k}(t, \mathbf{x}; u_k)\right) + \mathbf{D}(t, \mathbf{x}; u_k) : \mathcal{H}_\mathbf{x}\left(p_{\mathbf{X}|u_k}(t, \mathbf{x}; u_k)\right),$$
$$t \in (\tau_k, \tau_k + \hbar]$$

7.     Formulate MPC cost

$$\mathcal{J}(u_k) = \int_{\tau_k}^{\tau_k+\hbar} \int_{\Omega_\mathbf{x}} \left(\frac{1}{2}\left(p_{\mathbf{X}|u_k}(t, \mathbf{x}; u_k) - p_\mathbf{X}^d(t, \mathbf{x})\right)^2 + \frac{c}{2}u_k^2\right) d\mathbf{x}\, dt$$

8.     Solve KKT system with AD to obtain $u^\star$

$$\begin{bmatrix} \frac{\partial^2 \mathcal{L}}{\partial u_k^2} & \frac{\partial \mathcal{D}_k p_{\mathbf{X}|u_k}}{\partial u_k} \\ \frac{\partial \mathcal{D}_k p_{\mathbf{X}|u_k}}{\partial u_k} & 0 \end{bmatrix} \begin{bmatrix} \Delta u_k \\ \Delta \lambda \end{bmatrix} = -\begin{bmatrix} \frac{\partial \mathcal{L}}{\partial u_k} \\ \mathcal{D}_k p_{\mathbf{X}|u_k} \end{bmatrix}$$

9.     Apply $u = \text{clip}(u^\star, -50, 50)$, map to compliant chord
10.     Propagate aeroelastic and induced flow dynamics
11.     $t \leftarrow t + \Delta t$, $u_{-1} = u$
12. **end while**

## 4. The Deep Learning Approach for Density Evolution

To solve the probability density evolution problem, deep learning techniques are incorporated. This subsection details the rationale for using deep neural networks (DNNs) and outlines their implementation. It has been established that DNN models possess the capability to serve as universal function and operator approximators [40, 41], meaning that they can be used to approximate any nonlinear mapping, regardless of its complexity [42]. This efficacy of DNN models has been demonstrated in various disciplines from science, engineering, and applied mathematics [43]. A DNN is a finite composition of elementary operations. Hence, the chain rule of differentiation and the Automatic Differentiation (AD) concept can be used to find the derivatives of its outputs with respect to the inputs or any parameters in the network model. AD can be used to find these derivatives in either forward or backward paths without facing round-off or truncation errors [44]. In a Physics-Informed Neural Network (PINN) [45], this idea is used to encode differential operators within a DNN[42]. In these network models, the loss function for a PINN is defined such that it represents a differential equation in a residual form[45].

## 4.1 Physics-Informed Learning Frameworks: DeepPDEM and FPK-DP Net

In a recent study, the DeepPDEM network was proposed[23]. DeepPDEM is a DNN framework that utilizes PINN concept to encode the General Density Evolution Equation (GDEE) [46] into the network. Utilizing the equivalent dimension-reduced flux form of the FPK equation, DeepPDEM was later used to develop FPK-DP Net[22]. FPK-DP Net solves the FPK equation which gives the density evolution governed by Eq. (33). DNNs usually demand a large quantity of labeled data for training that may not be available or be expensive to compile. However, FPK-DP Net does not rely on any labeled training data. FPK-DP Net learns the FPK equation through randomly selected points from the domain and minimizing the differential equation in its residual form. In this research, the FPK-DP Net is generalized to incorporate the control input and is used to solve the density-shaping control problem.

Let $\mathbf{h}(t, \boldsymbol{\Theta})$ denote the decomposition of the stochastic input $\mathbf{W}_t$ using the Kosambi–Karhunen–Loève (KKL) theorem or Stochastic Harmonic Function (SHF) representation [47]. Thus, the stochastic differential equation given in Eq. (31) over the prediction horizon can be rewritten as [48],

$$\dot{\mathbf{X}}_t = \mathbf{f}(t, \mathbf{X}_t; u_k) + \mathbf{g}(t, \mathbf{X}_t; u_k)\mathbf{h}(t, \boldsymbol{\Theta}) \quad t \in (\tau_k, \tau_k + \hbar] \tag{47}$$

where $\mathbf{h}(\cdot)$ is a real-valued continuous function and $\boldsymbol{\Theta}$ is an uncorrelated random variable vector with support $\Omega_\Theta$ and joint PDF $p_{\boldsymbol{\Theta}}(\boldsymbol{\theta})$. Using this approximation, studies have shown that the probability density for each state $X_i$ can be approximated by the Generalized Density Evolution Equation (GDEE) [48].

$$\partial_t \tilde{p}_{X_i \Theta | u_k}(t, x_i, \boldsymbol{\theta}; u_k) + \dot{X}_i(t, \boldsymbol{\theta}; u_k) \partial_{x_i} \tilde{p}_{X_i \Theta | u_k}(t, x_i, \boldsymbol{\theta}; u_k) = 0 \tag{48}$$

Solving (48) gives $\tilde{p}_{X_i \Theta}(t, x_i, \boldsymbol{\theta}; u_k)$, which can be integrated over $\Omega_\Theta$ to find $\tilde{p}_{X_i}(t, x_i; u_k)$ [48].

$$\tilde{p}_{X_i | u_k}(t, x_i; u_k) = \int_{\Omega_\Theta} \tilde{p}_{X_i \Theta | u_k}(t, x_i, \boldsymbol{\theta}; u_k) d\boldsymbol{\theta} \tag{49}$$

Both the GDEE and the FPK equations represent the same density evolution. However, previous studies show that solving the FPK equation provides better results in terms of accuracy[49]. On the other hand, the marginal PDF found by GDEE can be used to reformulate the dimension-reduced equation into an equivalent decoupled

FPK-like PDE that can be effectively solved using a physics-informed DNN architecture. Given the solution of the GDEE, let $\tilde{f}_i(t, x_i)$ denote the equivalent drift coefficient of the FPK-like PDE governing the probability density evolution of $x_i$ [22].

$$\tilde{f}_i(t, x_i; u_k) = \frac{\int_{\Omega_\Theta} \left( \dot{X}_i(t, \boldsymbol{\theta}; u_k) - \sum_{j=1}^{m} \left( g_{ij}(t, X_i(t, \boldsymbol{\theta}; u_k)) h_j(t, \boldsymbol{\theta}) \right) \right) \tilde{p}_{X_i \boldsymbol{\Theta} | u_k}(t, x_i, \boldsymbol{\theta}; u_k) d\boldsymbol{\theta}}{\int_{\Omega_\Theta} \tilde{p}_{X_i \boldsymbol{\Theta} | u_k}(t, x_i, \boldsymbol{\theta}; u_k) d\boldsymbol{\theta}} \tag{50}$$

Which yields the equivalent decoupled form of the FPK equation as [22],

$$\frac{\partial \hat{p}_{X_i | u_k}(x_i, t; u_k)}{\partial t} = -\frac{\partial}{\partial x_i} \left( \tilde{f}_i(x_i, t; u_k) \hat{p}_{X_i | u_k}(x_i, t; u_k) \right) + D_{ii} \frac{\partial^2 \hat{p}_{X_i | u_k}(x_i, t; u_k)}{\partial x_i^2} \tag{51}$$

## 4.2 FPK-DP Net Architecture

In this section the architecture of the FPK DP Net is explained. Let $\tilde{p}_{X_i \boldsymbol{\Theta}}(t, x_i, \boldsymbol{\theta}; u_k)$ denote a composition function that incorporates deep networks $\mathcal{N}_G(t, x_i, \boldsymbol{\theta}, u_k; \boldsymbol{\chi}_G)$ to approximate the instantaneous solution to GDEE, $p_{X_i}(t, x_i, \boldsymbol{\theta}; u_k)$. The Initial Condition (IC) and the Boundary Condition (BC) can be hard-coded into the network model by including them explicitly in the composition function $\tilde{p}_{X_i \boldsymbol{\Theta}}(x_i, \boldsymbol{\theta}, t)$. Where $p_0(\cdot)$, $\mathcal{N}_{G_1}(\cdot)$, and $\mathcal{N}_{G_2}(\cdot)$ represent the initial joint pdf (IC), a strictly positive bounded output network, and an unbounded output network, respectively. To ensure the output is strictly positive, $\mathcal{N}_{G_1}(\cdot)$ utilizes a ReLU (Rectified Linear Unit – $\max(0, \mathbf{x})$) output layer. A scaled exponential $(\text{sgn}(\mathbf{x})(\exp(|\mathbf{x}|) - 1))$ output layer is used for $\mathcal{N}_{G_2}(\cdot)$ to guarantee the output unboundedness and the vanishing density boundary condition. Assuming that there exists a bounded uniformly Lipchitz solution, it has been shown that DeepPDEM network model can be used to find the solution to GDEE [23].

## 4.3 Training Strategy and Hyperparameters

Let "$\mathcal{L}_G$" be the approximation error (loss function) in $L^2$ sense. Since the boundary and initial conditions are already enforced, the loss function is used to learn the GDEE differential operator.

$$\mathcal{L}_G(\tilde{p}_{X_i \boldsymbol{\Theta} | u_k}) = \alpha_1 \left\| \partial_t \tilde{p}_{X_i \boldsymbol{\Theta} | u_k}(x_i, \boldsymbol{\theta}, t; u_k, \boldsymbol{\chi}_G) + \dot{X}_i(\boldsymbol{\theta}, t; u_k) \partial_{x_i} \tilde{p}_{X_i \boldsymbol{\Theta} | u_k}(x_i, \boldsymbol{\theta}, t; u_k, \boldsymbol{\chi}_G) \right\|_2^2, \quad \boldsymbol{\theta} \in \Omega_{\boldsymbol{\theta}} \tag{52}$$

where $\alpha_1$ is a strictly positive normalizing coefficient. The loss function can be approximated by $s_G$ collocation points that are chosen sufficiently large. Denoting the $j^{th}$ sample point as the superscript and the initial loss of the network as $\hat{\mathcal{L}}_G^0$, the approximated loss function is,

$$\hat{\mathcal{L}}_G(\tilde{p}_{X_i\Theta|u_k})$$
$$= \frac{1}{s_G \hat{\mathcal{L}}_G^0} \sum_{j=1}^{s_G} \left\| \partial_t \tilde{p}_{X_i\Theta|u_k}(x_i^j, \Theta^j, t^j; u_k, \chi_G) \right. \quad (53)$$
$$\left. + \dot{X}_i(\Theta^j, t^j; u_k) \partial_{x_i} \tilde{p}_{X_i\Theta|u_k}(x_i^j, \Theta^j, t^j; u_k, \chi_G) \right\|_2^2, \quad \Theta \in \Omega_\Theta$$

Hence, the network training can be denoted as the following optimization problem,

$$\hat{\mathcal{L}}_G^\star(\tilde{p}_{X_i\Theta|u_k}; \chi_G^\star) = \min_\chi \hat{\mathcal{L}}_G(\tilde{p}_{X_i\Theta|u_k}; \chi_G) \quad (54)$$

The training sequence of the DeepPDEM network is shown in Figure 2.

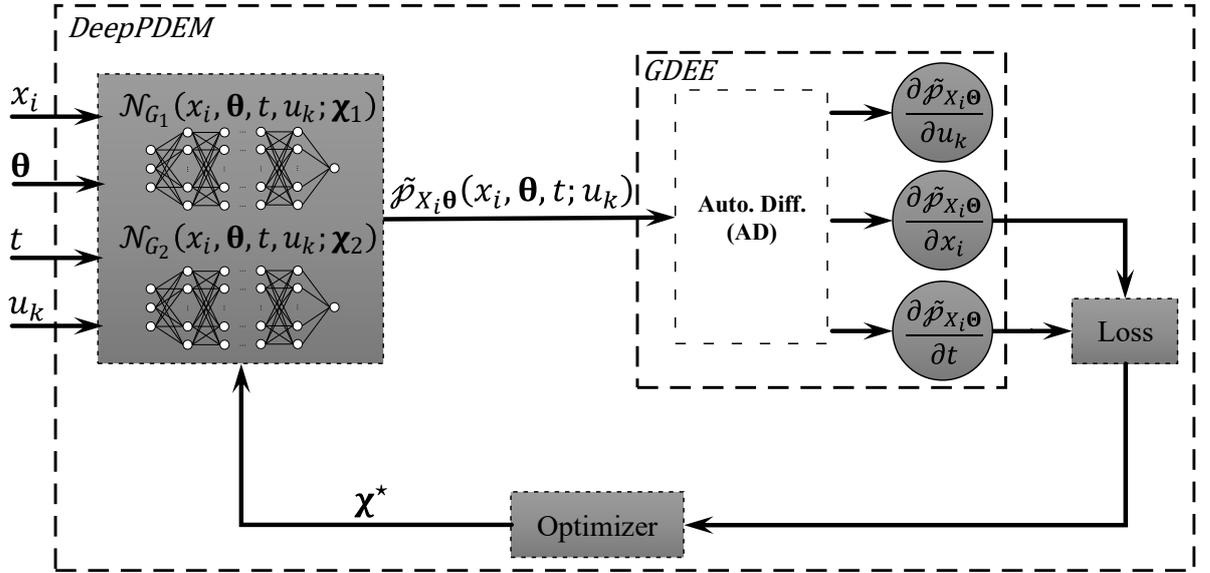

Figure 2 Schematic workflow representing the DeepPDEM network model training sequence

The output of the DeepPDEM network is then used to find the equivalent drift coefficient $\tilde{f}_i(t, x_i; u_k)$, as defined in Eq. (50). The equivalent drift coefficient is then input to another network model. Let $\hat{p}_{X_i}(t, x_i)$ denote a composition function that uses deep networks $\mathcal{N}_{F_1}(t, x_i, u_k; \chi_F)$ and $\mathcal{N}_{F_2}(t, x_i, u_k; \chi_F)$ to approximate the instantaneous solution of Eq. (51).

$$\hat{p}_{X_i}(t, x_i; u_k) = \mathcal{N}_{F_1}(t, x_i, u_k) \cdot \exp\left(-\left(\sqrt{-\ln p_0(x_i)} + t\mathcal{N}_{F_2}(t, x_i, u_k)\right)^2\right) \tag{55}$$

Now let "$\mathcal{L}_F$" and "$s_F$" be the approximation error (loss function) in $L^2$ sense and a set of collocation points, respectively.

$$\begin{aligned}
&\hat{\mathcal{L}}_F\left(\hat{p}_{X_i|u_k}\right) \\
&= \frac{1}{s_F \hat{\mathcal{L}}_F^0} \sum_{j=1}^{s_F} \left\| \frac{\partial}{\partial t} \hat{p}_{X_i|u_k}(x_i^j, t^j; u_k) + \frac{\partial}{\partial x_i} \tilde{f}_i(x_i^j, t^j; u_k) \hat{p}_{X_i|u_k}(x_i^j, t^j; u_k) \right. \\
&\quad \left. + \frac{\partial}{\partial x_i} \hat{p}_{X_i|u_k}(x_i^j, t^j; u_k) \tilde{f}_i(x_i^j, t^j u_k) - D_{ii} \frac{\partial^2}{\partial x_i^2} \hat{p}_{X_i|u_k}(x_i^j, t^j; u_k) \right\|_2^2
\end{aligned} \tag{56}$$

Once again, the superscript indicates the sample point counter and $\hat{\mathcal{L}}_F^0$ is the initial loss of the network. The training sequence of the DeepPDEM network is depicted in Figure 3.

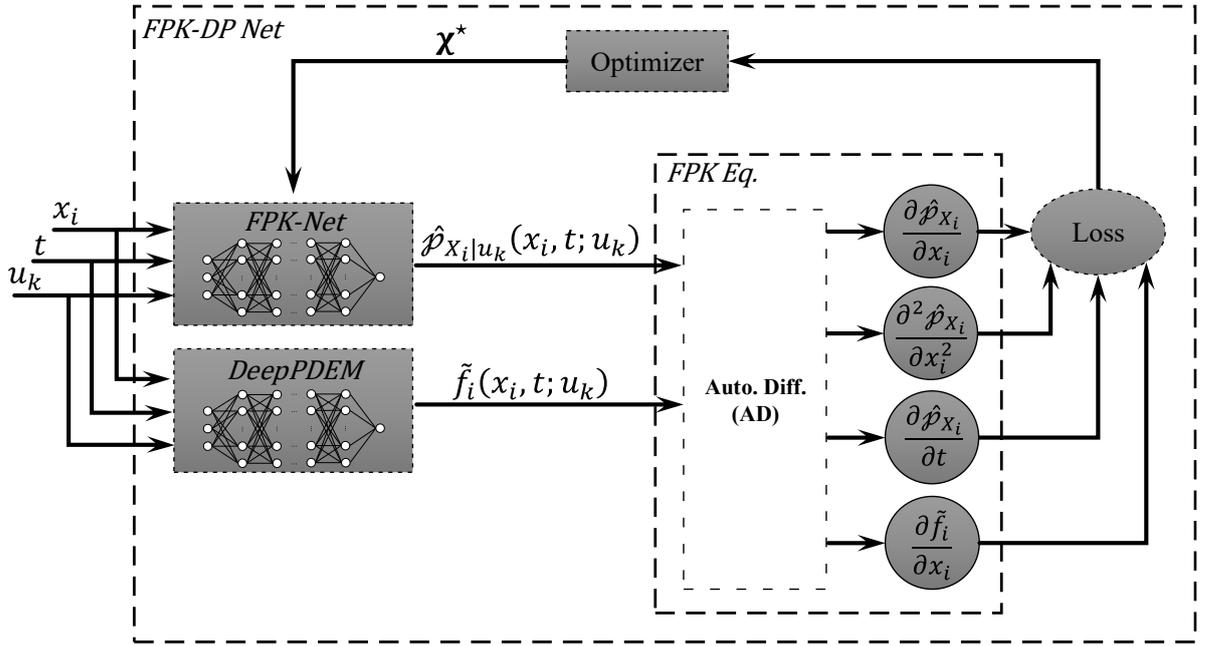

Figure 3 Schematic workflow representing the FPK-DP Net model training sequence

Based on previous literature studies, it is known that Multi-Layer Perceptrons (MLPs) can solve PDEs of the probability density evolution[45, 50, 51]. On the other hand, determining the appropriate number of neurons can be challenging, as it may vary in different problems. In this study, a network architecture with four layers and 20 neurons in each layer was chosen using trial and error. Each network is initialized using the Glorot normal initializer.

Gradient-based optimizers are the only feasible option in DNNs with numerous trainable variables. However, these optimizers may converge to local optima. To mitigate this issue, the Adam optimizer with an adaptive learning rate scheme is employed. Initiating the learning rate at a small value and gradually increasing it over the first few epochs of training helps achieve a balance between exploration and exploitation in the solution space. The learning rate is a critical parameter in the training process and significantly impacts the training and convergence rate. In this study, the rectified adaptive learning rate is used to update the learning rate during training. The initial value was set to the maximum possible value (0.015) which prevented the divergence of any network.

**Computational Environment and Run-Times**

The network models were implemented in TensorFlow, leveraging its built-in AD capabilities to compute the required partial derivatives. The Adam optimizer, which relies solely on first-order derivatives of the loss function, was employed with a rectified adaptive learning rate strategy, starting at 0.015 to ensure stable training. The number of layers and neurons per layer was selected empirically through extensive trial-and-error trade-offs, acknowledging that there is currently no universal guideline for optimal architecture design. Deeper networks generally provide higher representational capacity but at the cost of increased computational complexity and significantly reduced convergence rate. For the present proof-of-concept study, a four-layer architecture with 20 neurons per layer was adopted, as detailed earlier. A more detailed description of the training and tuning procedures for the network models can be found in [22] and [23].

Training times were observed to grow rapidly as the spatiotemporal domain increased, particularly for long-time integrations. While the deep neural networks employed here are inherently parallelizable, physics-constrained training remains computationally more expensive than traditional solvers such as finite element methods. In this study, the FPK-DP Net required approximately 54,395 s of wall-clock time to train on a legacy GPU. However, once trained, the network delivers predictions with negligible overhead, requiring only about 10 ms per evaluation step, which makes it highly suitable for online deployment. It should be noted that the choice of effective architecture remains an open research problem. Methods such as meta-learning may provide a systematic means of optimizing architecture and addressing the training-time limitation in future studies.

Table 1 – Summary of the network architecture and training configuration

| Component | Configuration/Value |
|---|---|
| Network Type | Multi-Layer Perceptron (MLP) |
| Num. of Layers | 4 Hidden Layers |
| Neurons Per Hidden Layer | 20 |
| Activation Functions | $\mathcal{N}_{G_1}(\cdot) \to ReLU$ <br> $\mathcal{N}_{G_2}(\cdot) \to$ Scaled Exponential |
| Initialization | Golrot Normal Initializers |
| Optimizer | Adam |
| Learning Rate Scheme | Adaptive Rectified |
| Initial Learning Rate | 0.015 |

## 5. Numerical Studies and Analysis

This section presents a proof-of-concept implementation to demonstrate the utility of the proposed density shaping method for active CW gust load alleviation control via numerical simulations. The case study considers an unswept, untapered cantilever wing with uniform mass and a NACA0012 section, operating below flutter speed under moderate gusts. The proposed workflow is generally applicable to other flexible wings, as it relies on the probability density evolution rather than statistical moments. However, quantitative performance (RMS reductions, required control effort, power-difference gains) will not transfer without re-modeling whenever wing planform, mass distribution, or flight regime changes substantially. Similarly, this also applies to conditions characterized by very large angles of attack or post-stall flows, as well as very strong and impulsive turbulence intensities.

As for the practical implications of this study, when considering new wing geometries, a new modal basis and re-identification of the reduced aeroelastic model are required. Furthermore, higher speeds or separated flows necessitate higher-fidelity unsteady aerodynamics or corrected diffusion/drift estimates prior to solving the FPK equation. Additionally, high turbulence or actuator limits may require constraint changes in the MPC and may degrade closed-loop gains.

First, the numerical simulation, its results, and validation are reviewed. Subsequently, the open loop response of the CW is investigated under two stochastic excitations: the Band-Limited White Noise and the Dryden stochastic clear-air turbulence (CAT) model. Finally, the performance of the proposed control methodology for alleviating stochastic gust loads is demonstrated.

The aero-structural model used in this research is validated by investigating the flutter boundary of the Goland wing as shown in Figure 4. The predicted flutter velocity and frequency were $135\ m/s$ and $69\ rad/s$, respectively, which are in good agreement with the results presented in [52].

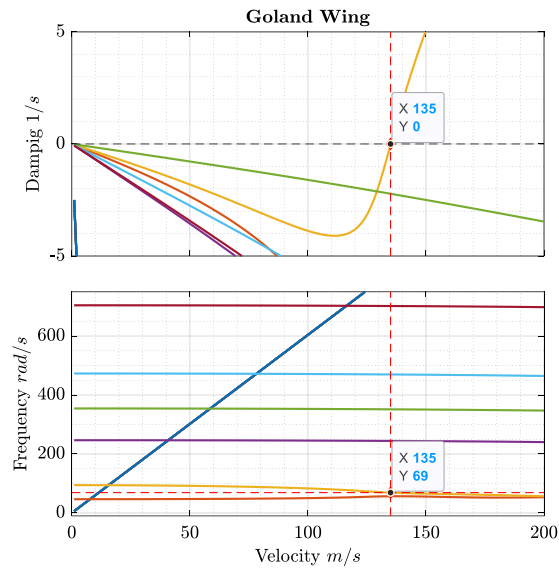

Figure 4 - Validation of the aero-structural model through flutter boundary prediction of the Goland wing, where the flutter boundaries are obtained by coupling the structural equations of motion with Peters' finite-state aerodynamic model.

The compliant chord model was also validated by analyzing the 2-DoF model properties and simulation results presented in [28] using the unsteady aerodynamics. The ratio of the morphing segment ($\ell_c/2b$) and the relative stiffness, were set to 25% and 50%, respectively. The predicted flutter velocity and frequency are found to be $42.03\ m/s$ and $7.18\ Hz$. These results are in good agreement with the values presented in [28] ($41.7\ m/s$ and $7.2\ Hz$), indicating the validity of the implementation process.

Table 2 provides a comprehensive list of properties for a case-study wing selected for the numerical investigations based on the wing studied in[53]. After extensive simulations, ten structural modes and six aerodynamic states were found to be sufficient for simulating the behavior of the CW. The ten modes consist of five bending modes and five torsional modes. A flutter analysis is essential to sufficiently assess the aeroelastic behavior of the case-study wing. Figure 5 and Figure 6 illustrate the flutter boundaries and the static response of the wing at $\alpha = 10°$, respectively.

Table 2 - Baseline aerodynamic and structural properties of the case-study compliant wing (CW)

| Parameter | Value |
|---|---|
| Wing Half Span ($m$) | 0.8 |
| Root Chord ($m$) | 0.3 |
| Taper Ratio | 0 |

| Parameter | Value |
| --- | --- |
| Mass ($kg$) | 0.849 |
| Bending Rigidity ($N.m^2$) | 740.2947 |
| Torsional Rigidity ($N.m^2$) | 438.2651 |
| Airfoil | NACA0012 |
| Elastic Axis, from LE (m) | 0.15 |
| C.G. Loc., from LE (m) | 0.15 |
| Mass per unit length ($kg/m$) | 1.0612 |
| Structural Density ($kg/m^3$) | $3.2043e3$ |
| Mass Moment of Inertia ($kg/m^2$) | diag($[0.045, 1.1878e-5, 9.6875e-5]$) |
| Free Stream Vel. ($m/s$) | 25 |
| Free Stream Dens. ($kg/m^3$) | 1.225 |
| Compliant Chord (%) | 50 |
| Num. of Structural Modes | 5 (bending), 5 (torsional) |
| Num. of Aerodynamic States | 6 |

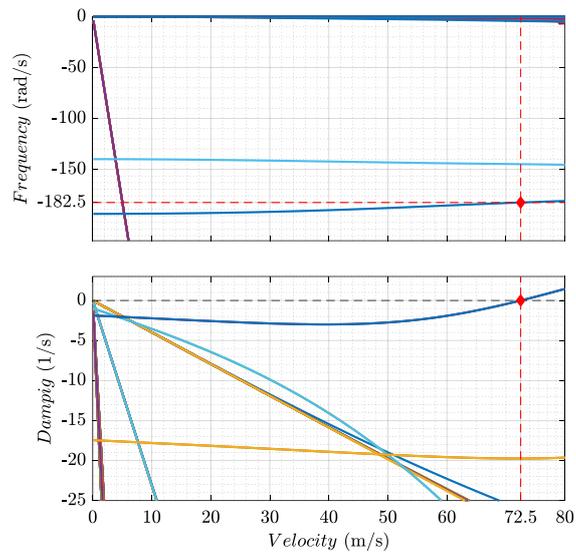

Figure 5 - Flutter boundaries of the case-study wing, illustrating the flutter boundaries obtained from the coupled aeroelastic model of the compliant wing using Peters' finite-state aerodynamic formulation

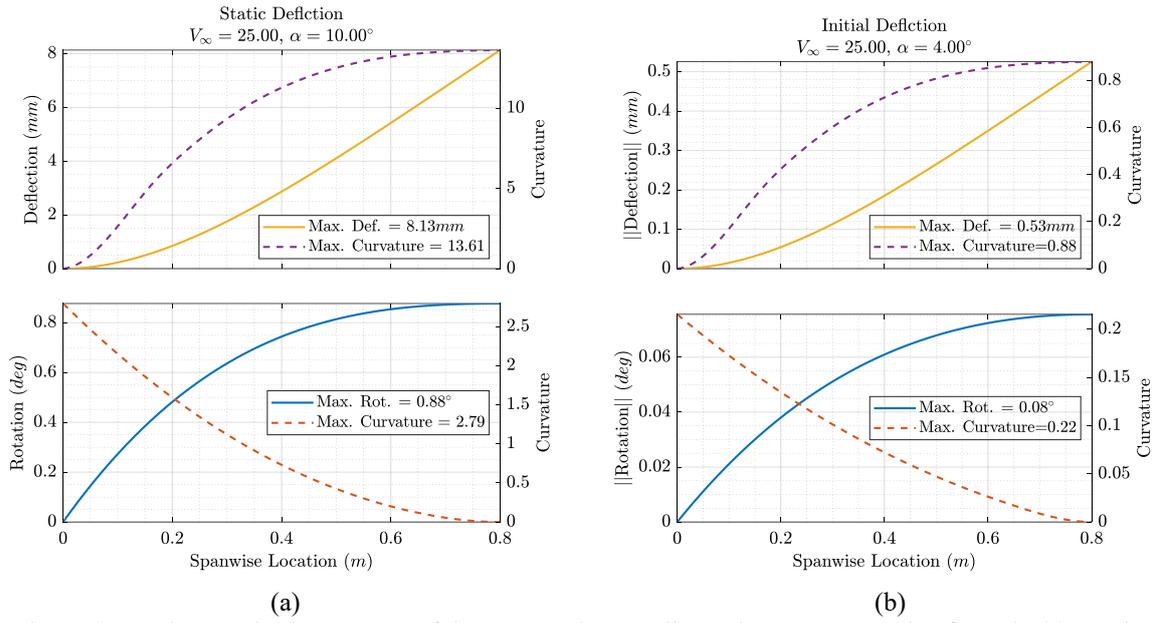

Figure 6 - Static aeroelastic response of the case-study compliant wing at a 10° angle of attack. (a) Vertical deflection distribution along the wingspan. (b) Wing twist distribution along the span.

Subsequently, as the focus of the present study is to alleviate loads and control the wing vibrations under stochastic gust excitations, the free stream velocity is chosen below the flutter velocity. As indicated before, to examine the wing behavior under disturbances, two different stochastic excitations, i.e., Band-Limited White Noise (BLWN) and Dryden stochastic turbulence model, are utilized. The vertical gust velocity strength ($3\sigma$) for both signals is assumed to be equal to 5 $m/s$ i.e. and the gust induced AoA strength to be $\pm 10°$. Furthermore, the wing is assumed to be initially in a steady flight condition, encountering the gusts at $t = 0$ and remaining in the gust field for the entire simulation time. Moreover, the study presumes a wing root angle of attack of 4 degrees and presents all deflections relative to this state. Additionally, the maximum control moment input is limited to 50 $N.m$. Figure 7 (a) and (b) represent sample profiles of the stochastic vertical gust and their Power Spectral Density, respectively. In addition, Table 3 provides an overview of the parameters and the simulation environment.

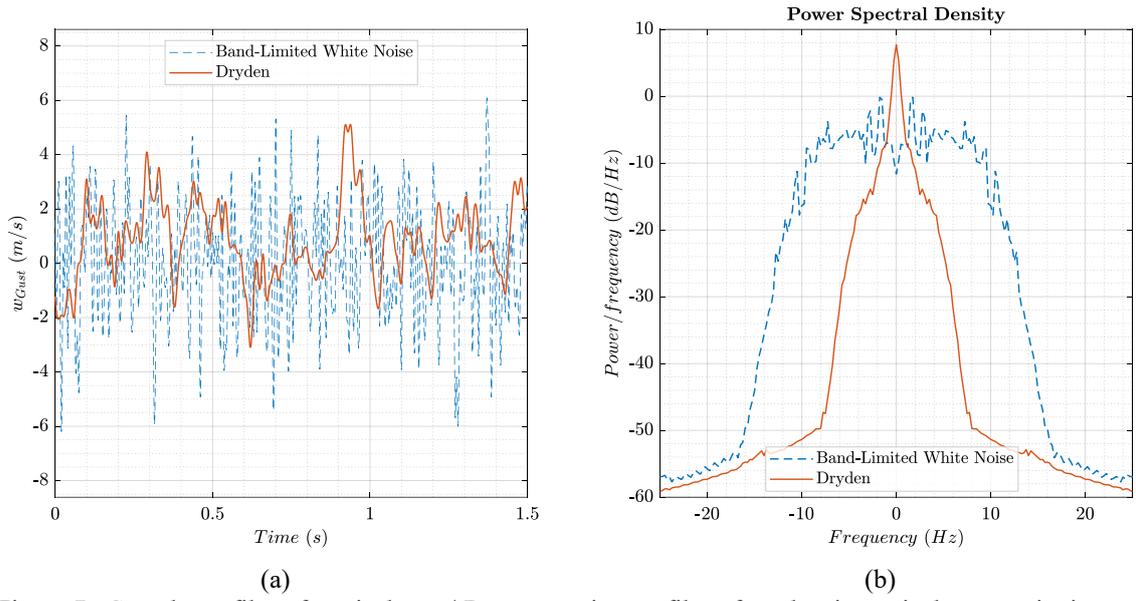

Figure 7 - Sample profiles of vertical gust | Representative profiles of stochastic vertical gust excitations. (a) Time history of the vertical gust velocity for both the Band-Limited White Noise (BLWN) and Dryden turbulence models. (b) Corresponding power spectral density (PSD) of the gust signals.

Table 3 - Summary of gust excitation, operating, and numerical simulation parameters

| Gust Model | BLWN/Dryden |
|---|---|
| Vertical Gust Velocity Strength ($3\sigma$) | $5 \, m/s$ |
| Gust-Induced AoA Perturbation | $\pm 10°$ |
| Wing-Root Trimmed AoA | $4°$ |
| Maximum Control Input Moment | $< 50 \, N \cdot m$ |
| Simulation Time Step | $0.01 \, s$ |

A time step of $0.01 \, s$ is used in all time domain simulations. The gust-induced vibration of the wing is first simulated with no actuation (in an open loop fashion). Figure 8 (a) and (b) illustrate the uncontrolled wing tip vertical deflection and twist over $10 \, s$ subjected to the two gust profiles, respectively. The vertical deflection Root Mean Squares (RMS) are $8.783 \, mm$ and $6.938 \, mm$ for the BLWN and Dryden gust profiles, respectively. The twist RMS values at the wing tip are also $0.566°$ and $0.485°$, respectively.

Figure 9 (a) displays the wing root bending moment and, (b) illustrates the correlation between wing bending and torsional moments. The root bending moment RMS values are $36.726 \, N.m$ and $30.939 \, N.m$ for the BLWN and Dryden gust profiles, respectively. The coefficient of variation ($\hat{c}_v$) for the bending and torsional moments are $2.2$ and $-2.3$ for the BLWN gust profile, and $2.2$ and $-2.4$ for the Dryden gust profile, respectively. In addition, Figure 10 shows the probability density distribution evolution of the wing tip vertical deflection, evaluated by the FPK-DP Net.

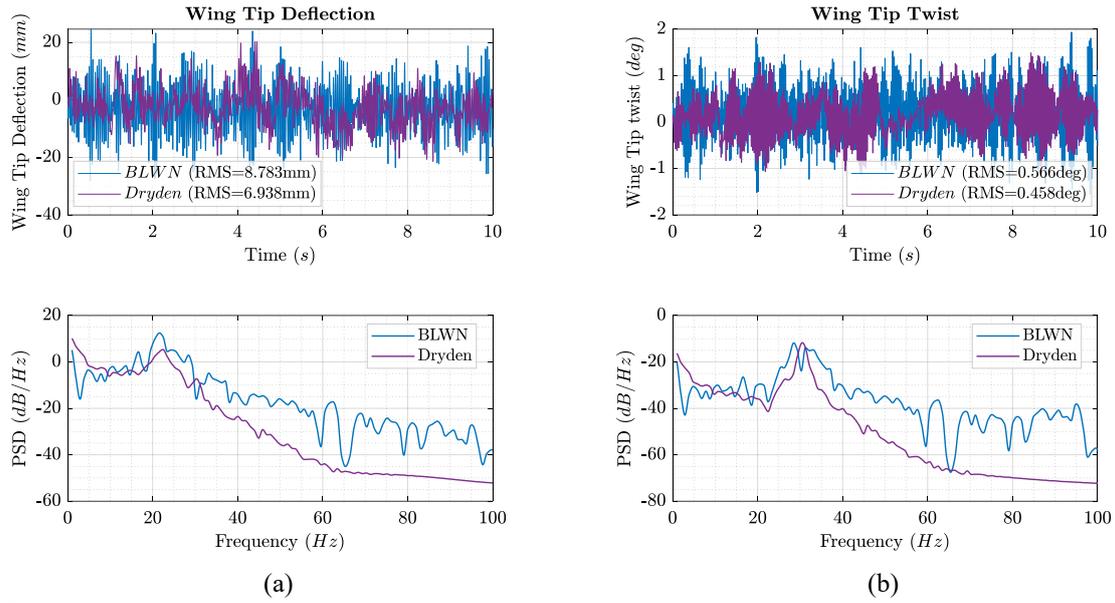

(a)                          (b)

Figure 8 – Open-loop response of the compliant wing tip subjected to stochastic gust excitations over a 10 s interval. (a) Vertical deflection and (b) twist of the wing tip under Band-Limited White Noise (BLWN) and Dryden turbulence profiles.

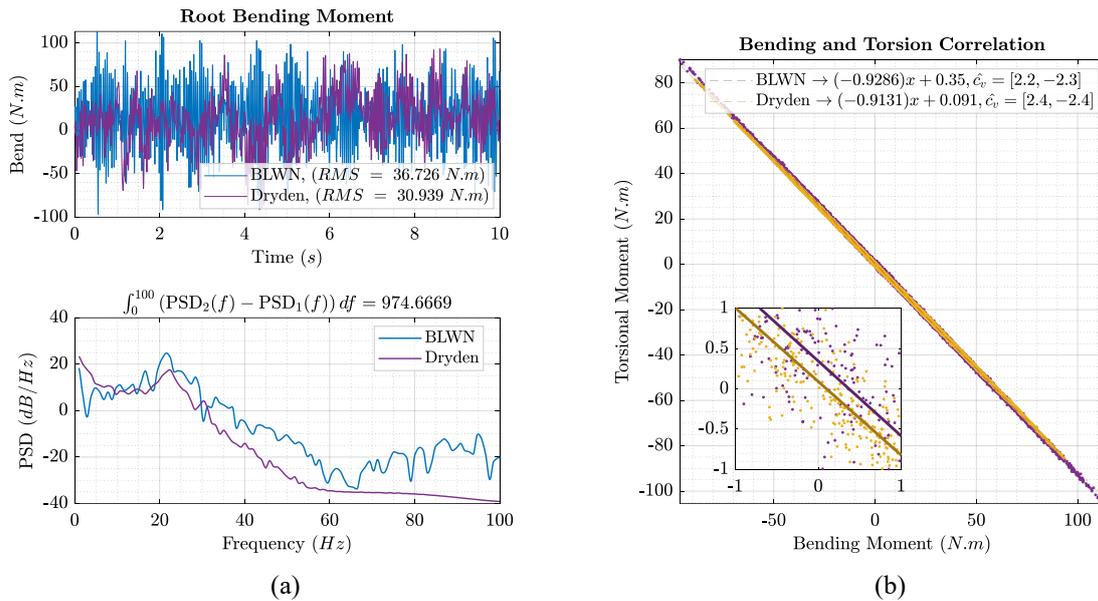

(a)                          (b)

Figure 9 – Open-loop response of the compliant wing subjected to stochastic gust excitations over a 10 s interval. (a) Root bending moment and (b) wing root bending-torsional moment correlation under Band-Limited White Noise (BLWN) and Dryden turbulence profiles.

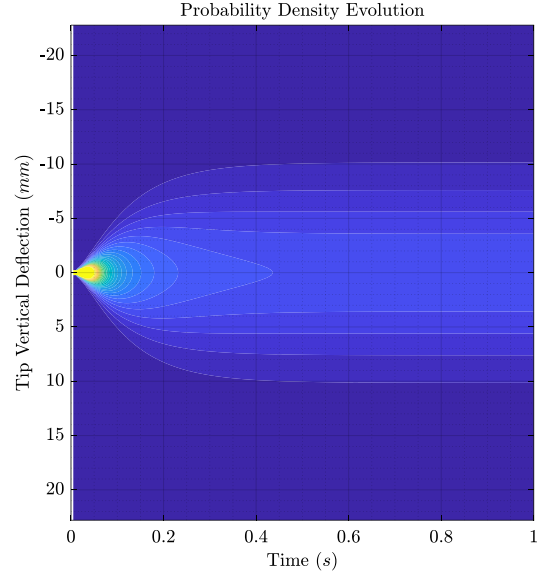

Figure 10 – Time evolution of the probability density distribution (PDF) of the wing tip vertical deflection under Band-Limited White Noise (BLWN) gust excitation, as predicted by the FPK-DP Net. The network captures the full stochastic response of the compliant wing in the absence of control actuation, illustrating the spread and evolution of the deflection distribution over the simulation horizon.

## 5.1 Density Shaping Control

This subsection evaluates the effectiveness of the proposed controller under different gust conditions by comparing controlled and uncontrolled responses. The effectiveness of the load alleviation is evaluated using both BLWN and Dryden gust profiles. The capability is assessed by comparing the wing tip vertical deflection, twist, root bending and torsional moments of the controlled wing with those of the unactuated wing. The load alleviation capability of the proposed scheme is first evaluated considering the Compliant Chord Fraction (CCF, $\ell_c/2b$) of 0.5. Next, the CCF is reduced from 0.5 to 0.3, decreasing the control authority, and the capability of the load alleviation scheme is re-evaluated.

Traditional gust control studies design controllers to suppress gust vibrations as much as possible. However, this approach can lead to a large control signal that could saturate the control input. To address this issue, the weighting parameter "$c$" in Eq. (38) should be carefully chosen or tuned. In the current study, the value of "$c$" has been chosen to be 0.1 via a trial-and-error process. This value results in an effective gust suppression while avoiding control saturation.

The proposed density shaping controller is applied to the wing model subjected to BLWN gust profile. Figure 11 (a) depicts the control signal acting on the compliant chord. The RMS value for the control signal is 21.1506 N.m. Additionally, Figure 11 (b) illustrates the evolution of the probability density calculated by the FPK-DP Net.

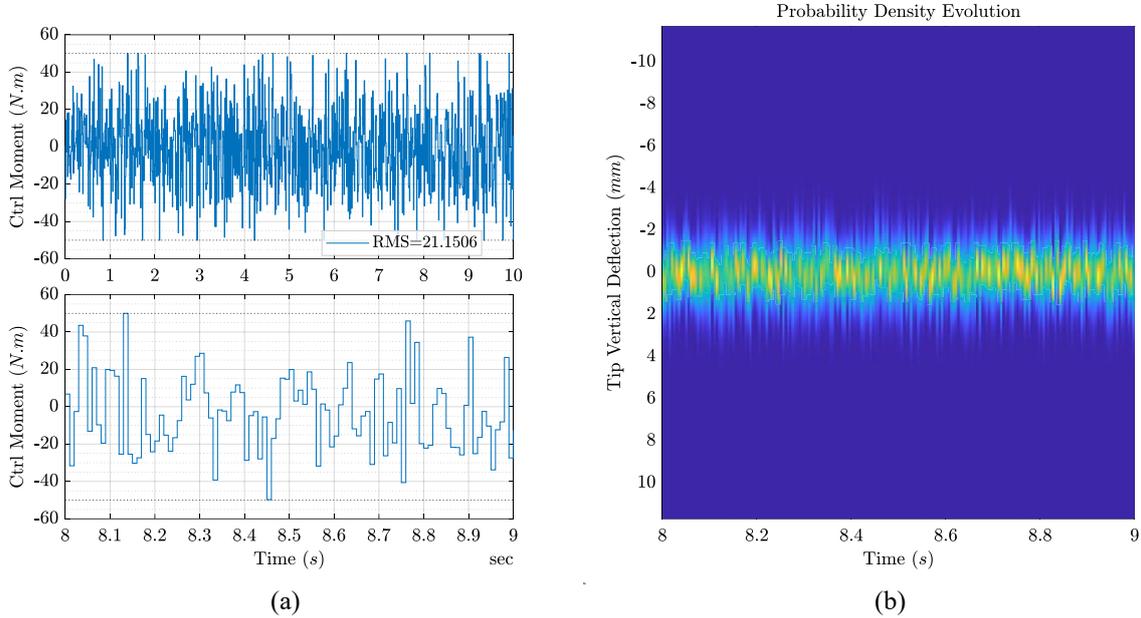

(a) (b)

Figure 11 – (a) Control signal applied to the compliant chord modulates wing camber to suppress stochastic gust-induced vibrations, as illustrated by (b) the wing tip vertical deflection temporal evolution of probability density distribution under the BLWN gust profile with a CCF of 0.5 in response to the applied control

The controlled wing tip vertical deflection and twist are compared with the open-loop response in Figure 12. The results indicate reduction of 30.24% and 43.19% in the wing tip vertical deflection and twist RMS values, respectively. To facilitate a more accurate comparison, a power difference ($\Delta_P$) as an oriented distance function is defined as:

$$\Delta_P = \int_0^{100} \left( \text{PSD}_{OL}(f) - \text{PSD}_{CL}(f) \right) df \tag{57}$$

Analyzing the power difference reveals that the closed-loop signals for the wing tip vertical deflection and twist have 621.7475 $dB$ and 610.8069 $dB$ less power, respectively, compared to the unactuated system. Moreover, as illustrated in Figure 13, the bending and torsional moment RMS values at the wing root are reduced by 21.21% and 19.55%, respectively. In addition, the power difference also reveals that the closed-loop signals for the wing root bending and torsional moments have 648.05 $dB$ and 631.94 $dB$ less power, respectively, compared to the open-loop system. The observed reduction in wing tip deflection, twist, and root moment is attributed to the active reshaping of the camber. The DSMPC utilizes the full probability density distribution information of the response, as computed by the FPK-DP Net, to synthesize the optimal control signal. This signal drives the compliant section to dynamically alter the wing's aerodynamic profile. By continuously adjusting the camber, the controller generates counteracting aerodynamic forces and moments that directly oppose and mitigate the effects of incoming stochastic gusts, thereby suppressing the aeroelastic vibrations and structural loads.

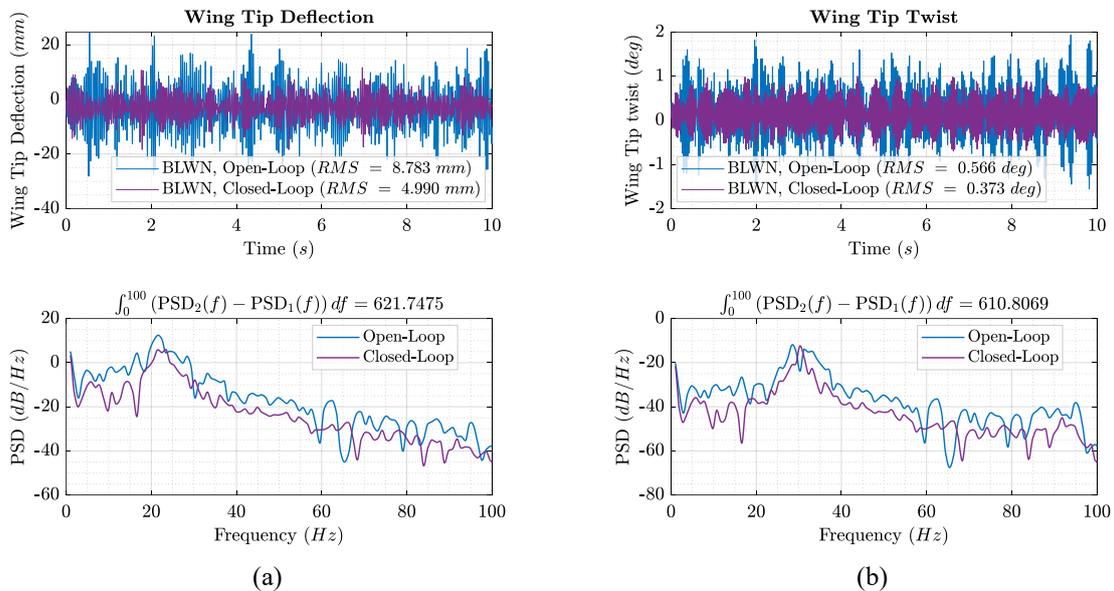

Figure 12 – Comparison of (a) the wing tip vertical deflection and (b) twist under BLWN gust excitation for a CCF of 0.5 demonstrating the effectiveness of the proposed density shaping controller in reducing wing tip deflection and twist RMS and power difference ($\Delta_P$) compared to the uncontrolled (open-loop) response.

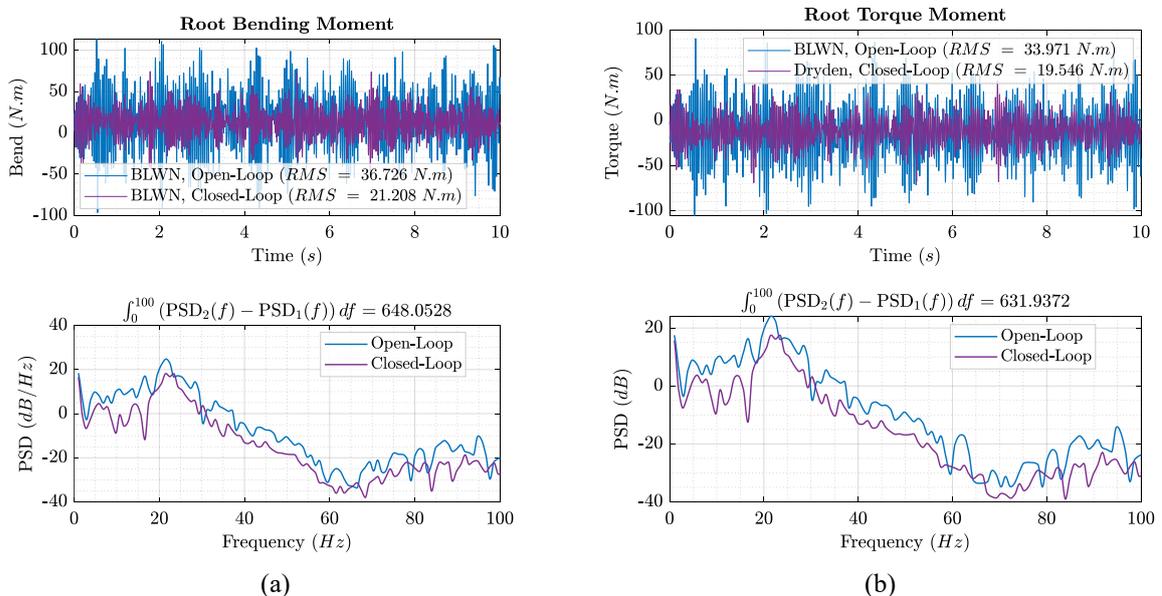

Figure 13 – Comparison of (a) the wing root bending and (b) torsional moments under BLWN gust excitation for a CCF of 0.5 demonstrating the effectiveness of the proposed density shaping controller in reducing wing root bending moment and torsion RMS and power difference ($\Delta_P$) compared to the uncontrolled (open-loop) response.

Next, the proposed density shaping controller is applied to the wing model subjected to the Dryden gust profile. Figure 14 (a) and (b) illustrate the control signal acting on the compliant chord and the evolution of the probability density distribution calculated by the FPK-DP Net, respectively. The RMS value for the control signal in this case is 21.7233 N.m.

Figure 15 compares the controlled versus unactuated responses of the wing tip vertical deflection and twist. As indicated in Figure 15, the wing tip vertical deflection and twist RMS values are reduced by 42.04% and 46.51%, respectively. Furthermore, the power differences reveal that the controlled wing tip vertical deflection and twist have 596.95 $dB$ and 683.54 $dB$ less power, respectively, compared to the unactuated system. Figure 16 depicts the wing root bending and torque moments RMS values, which are reduced by 41.16% and 41.35%, respectively. The power differences also indicate that the controlled wing root bending and torque moments have 575.46 $dB$ and 579.07 $dB$ less power, respectively, compared to the unactuated wing.

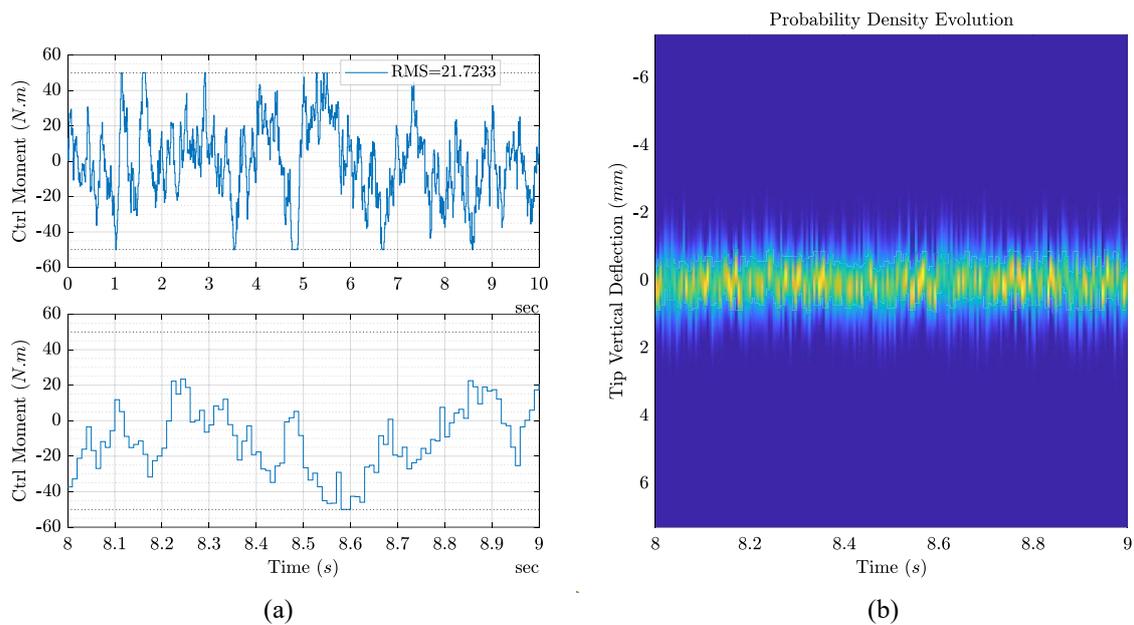

Figure 14 – (a) Control signal applied to the compliant chord modulates wing camber to suppress stochastic gust-induced vibrations, as illustrated by (b) the wing tip vertical deflection temporal evolution of probability density distribution under the Dryden gust profile with a CCF of 0.5 in response to the applied control

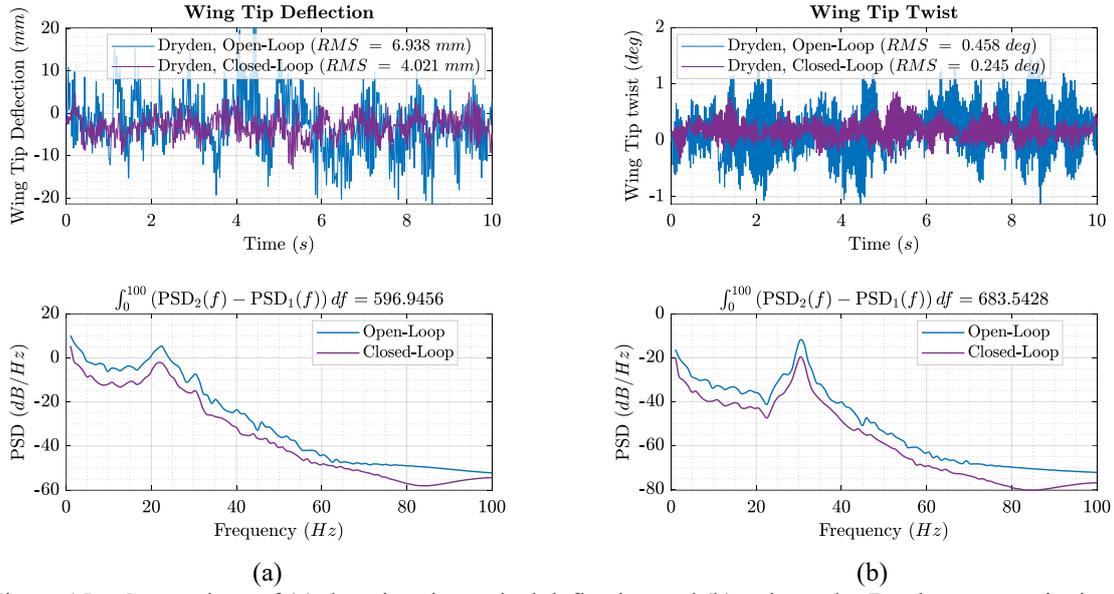

(a)          (b)

Figure 15 – Comparison of (a) the wing tip vertical deflection and (b) twist under Dryden gust excitation for a CCF of 0.5 demonstrating the effectiveness of the proposed density shaping controller in reducing wing tip deflection and twist RMS and power difference ($\Delta_P$) compared to the uncontrolled (open-loop) response.

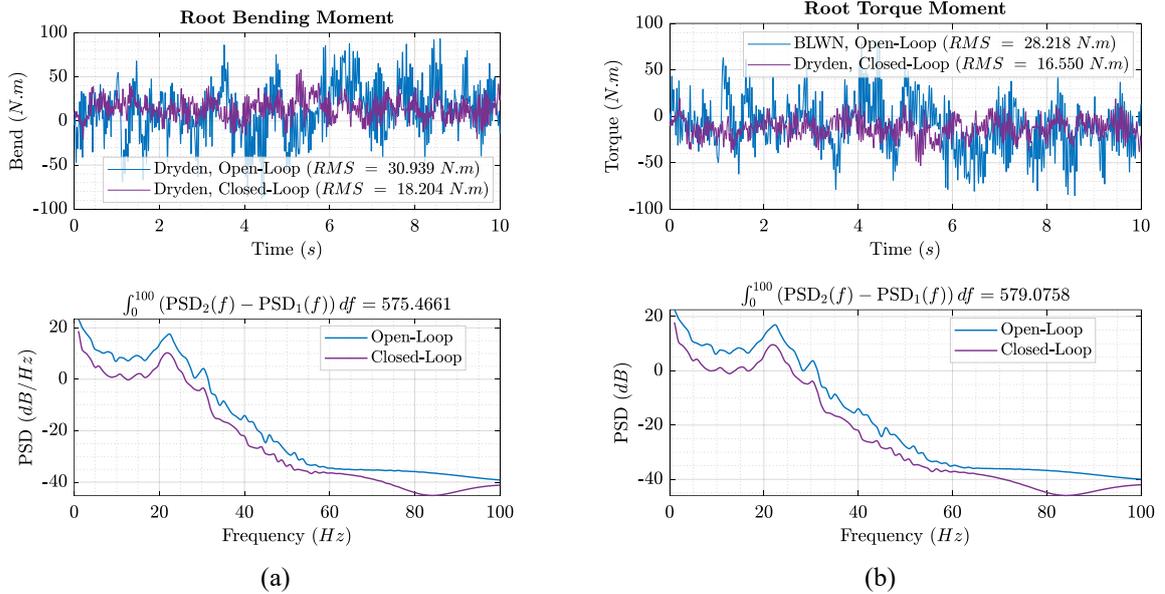

(a)          (b)

Figure 16 – Wing root bending and torsional moments, subjected to the Dryden gust profile and CCF of 0.5 | Comparison of (a) the wing root bending and (b) torsional moments under Dryden gust excitation for a CCF of 0.5 demonstrating the effectiveness of the proposed density shaping controller in reducing wing root bending moment and torsion RMS and power difference ($\Delta_P$) compared to the uncontrolled (open-loop) response.

Next, the study investigates the impact of reducing the CCF from 0.5 to 0.3 on the proposed density shaping controller. Despite the reduced CCF, the controller maintains its gust load alleviation capability and reduces wing tip deflections. However, the control signal RMS value increases by 43.34% due to decreased control authority. The wing tip vertical deflection and twist RMS values decrease by 30.24% and 43.19% respectively, compared to the uncontrolled wing, but increased by 18.42% and 130.03% respectively, compared to the CCF of 0.5. The wing root bending and torque moments RMS values decreased by 32.11% and 32.25% respectively, but increased by 17.57% and 17.74% respectively, compared to the CCF of 0.5.

When subjected to the Dryden gust profile, the wing model with a CCF of 0.3 shows an increase in the control signal RMS value by 38.30% compared to the CCF of 0.5. The wing tip vertical deflection and twist RMS values decreased by 28.81% and 38.21% respectively, but increased by 22.83% and 15.51% respectively, compared to the CCF of 0.5. The wing root bending and torque moments RMS values decreased by 28.37% and 28.41% respectively, but increased by 21.74% and 21.97% respectively, compared to the CCF of 0.5. Detailed results and corresponding graphs can be found in the appendix. The observation that a reduced CCF necessitates a higher control signal RMS value to achieve alleviation is physically intuitive. With a smaller compliant area available for active camber variation, a greater control effort (moment) is required to induce the necessary shape changes and generate sufficient counteracting aerodynamic forces to effectively suppress the gust-induced disturbances.

Table 4 and Table 5 provides a summary of the numerical studies' results presented in this research. As shown by the results, for both CCFs, the developed proof-of-concept implementation of the proposed density shaping gust load alleviation control maintains gust load alleviation capability and reduces the wing tip deflections.

Table 4 – Wing tip vertical deflection and twist | Comparison of the RMS values of wing tip vertical deflection and twist for both BLWN and Dryden turbulence profiles, evaluated at different CCFs

| Gust | CCF | Ctrl. Signal RMS ($N.m$) | Vertical Def. RMS ($mm$) | Twist RMS ($deg$) | Vertical Def. $\Delta_P$ | Twist $\Delta_P$ |
|---|---|---|---|---|---|---|
| BLWN | 0 | 0 | 8.783 | 0.566 | 0 | 0 |
| BLWN | 0.3 | 30.3179 | 5.909 | 0.858 | 323.50 | 289.3319 |
| BLWN | 0.5 | 21.1506 | 4.99 | 0.373 | 621.7475 | 610.8069 |
| Dryden | 0 | 0 | 6.938 | 0.458 | 0 | 0 |
| Dryden | 0.3 | 30.0425 | 4.939 | 0.283 | 368.32 | 515.884 |
| Dryden | 0.5 | 21.7233 | 4.021 | 0.245 | 596.95 | 683.5428 |

Table 5 – Comparison of root bending and torsional moment responses of the compliant wing subjected to BLWN and Dryden turbulence profiles for different CCF

| Gust | CCF | Controller Signal RMS ($N.m$) | Root Bending RMS ($N.m$) | Root Torque RMS ($N.m$) | Bending $\Delta_P$ | Torque $\Delta_P$ |
|---|---|---|---|---|---|---|
| BLWN | 0 | 0 | 36.726 | 33.971 | 0 | 0 |
| BLWN | 0.3 | 30.3179 | 24.935 | 23.014 | 353.079 | 327.6269 |
| BLWN | 0.5 | 21.1506 | 21.208 | 19.546 | 648.0528 | 631.9372 |
| Dryden | 0 | 0 | 30.939 | 28.218 | 0 | 0 |
| Dryden | 0.3 | 30.0425 | 22.162 | 20.186 | 351.2153 | 355.7433 |
| Dryden | 0.5 | 21.7233 | 18.204 | 16.55 | 575.4661 | 579.0758 |

# 6. Concluding Remarks

The proposed Density Shaping Model Predictive Controller (DSMPC) leverages a novel physics-informed machine learning framework, called FPK-DP Net, to compute the probability density distribution evolution of wing deflections in a Compliant Wing (CW) under stochastic gust excitations. By solving a dimension-reduced form of the Fokker-Planck-Kolmogorov (FPK) equation, the DSMPC utilizes the full probability density information, rather than relying solely on statistical moments such as mean or variance, to optimize wing camber variations for gust load alleviation and vibration suppression. In numerical evaluations, the DSMPC was integrated with structural equations of motion for the CW, coupled with a finite-state aerodynamic model validated against established literature. Under two representative stochastic gust profiles and for two Compliant Chord Fractions (CCF), the controller achieved approximately 30% reduction in Root-Mean-Square (RMS) wing tip deflection and 40% reduction in RMS twist, while also mitigating root bending and torsional loads.

The DSMPC framework demonstrates enhanced performance in nonlinear and uncertain flight environments by avoiding restrictive assumptions such as Gaussian distributions or linear system behaviors, thereby providing a richer and more accurate basis for control decisions. Unlike the existing stochastic control methods that act on statistical moments such as mean or variance, the proposed DSMPC uses the full probability density distribution information to shape the wing camber and alleviate stochastic loadings. Beyond its immediate technical contributions, this framework opens up new avenues for further innovative studies and a step forward in probabilistic control theory. This approach unifies stochastic uncertainty quantification with aero-servo-elastic control synthesis, enabling the derivation of advanced probabilistic metrics, including time-varying reliability indices.

While the present results are limited to numerical verification, the methodology facilitated by the proposed framework shows promise for application to morphing wing technologies and high-aspect-ratio aircraft, where active gust load alleviation and vibration suppression are critical for structural efficiency and operational safety. This combination of theory, computation, and demonstrated capability provides a scientifically grounded path toward more resilient and efficient active aeroelastic control and paves the way for and be potentially employed in service or predictive maintenance scheduling that foster safer and more economical aviation practices.

# References

# Appendix

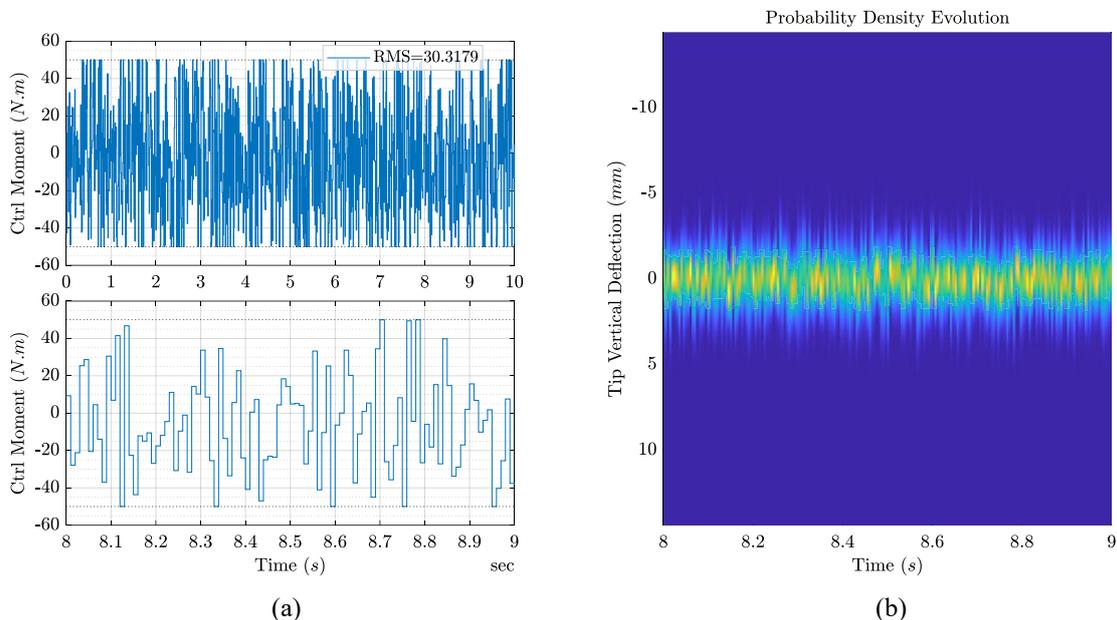

(a)        (b)

Figure 17 – (a) Control signal applied to the compliant chord modulates wing camber to suppress stochastic gust-induced vibrations, as illustrated by (b) the wing tip vertical deflection temporal evolution of probability density distribution under the BLWN gust profile with a CCF of 0.3 in response to the applied control

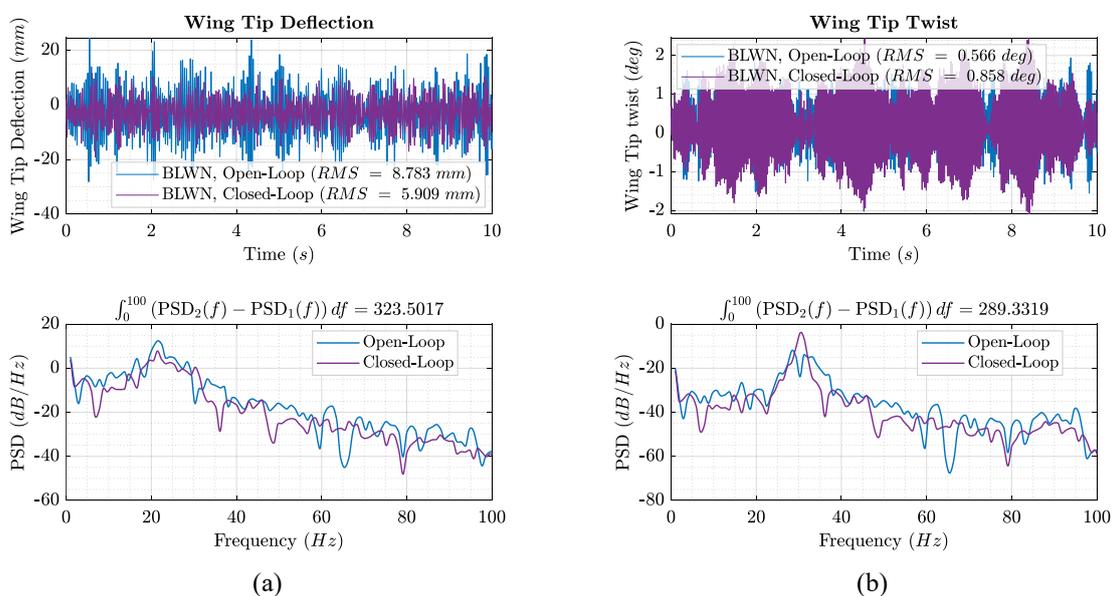

(a)        (b)

Figure 18 – Comparison of (a) the wing tip vertical deflection and (b) twist under BLWN gust excitation for a CCF of 0.3 demonstrating the effectiveness of the proposed density shaping controller in reducing wing tip deflection and twist RMS and power difference ($\Delta_P$) compared to the uncontrolled (open-loop) response.

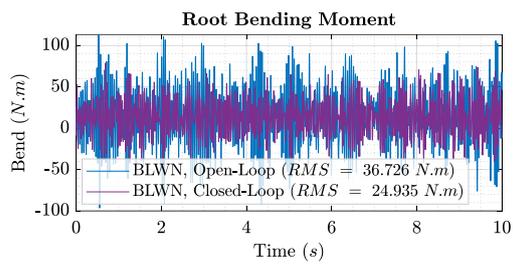
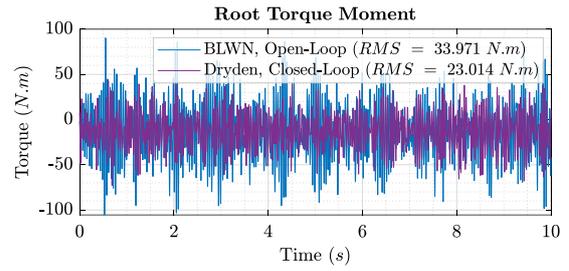
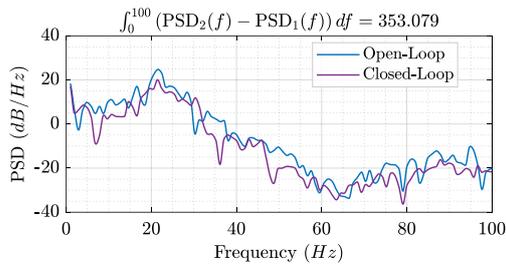
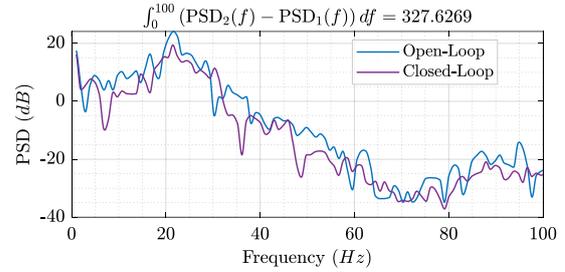

(a)        (b)

Figure 19 – Comparison of (a) the wing root bending and (b) torsional moments under BLWN gust excitation for a CCF of 0.3 demonstrating the effectiveness of the proposed density shaping controller in reducing wing root bending moment and torsion RMS and power difference ($\Delta_P$) compared to the uncontrolled (open-loop) response.

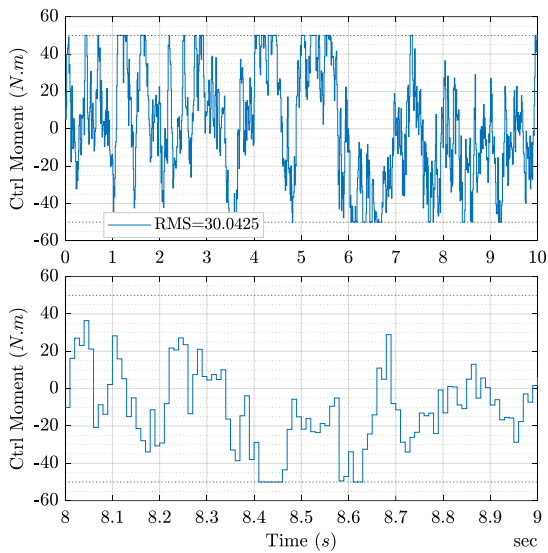
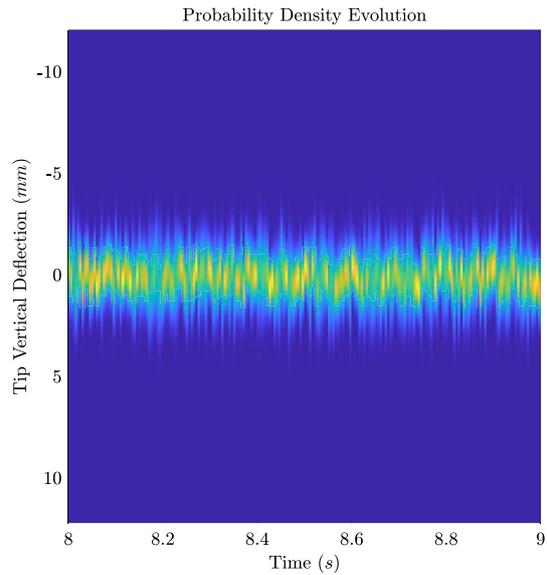

(a)        (b)

Figure 20 – (a) Control signal applied to the compliant chord modulates wing camber to suppress stochastic gust-induced vibrations, as illustrated by (b) the wing tip vertical deflection temporal evolution of probability density distribution under the Dryden gust profile with a CCF of 0.3 in response to the applied control

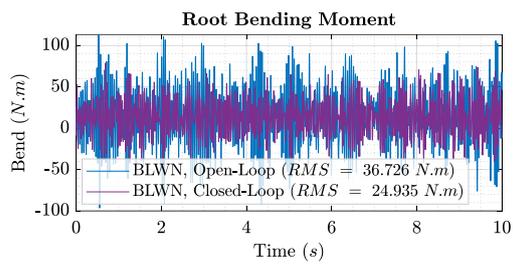
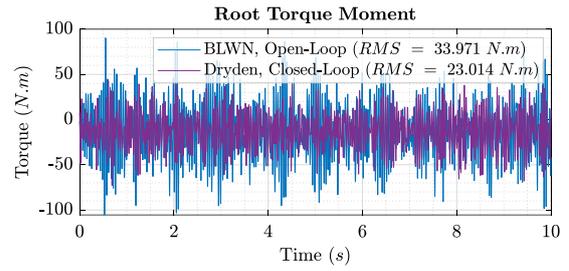
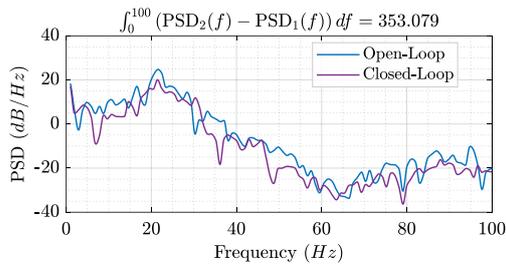
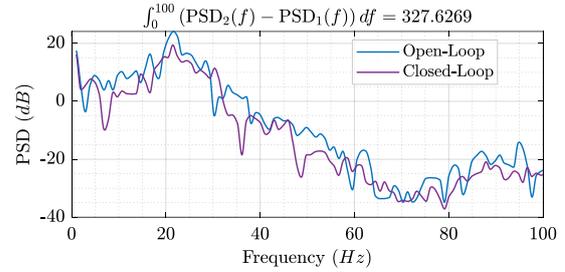

(a)        (b)

Figure 19 – Comparison of (a) the wing root bending and (b) torsional moments under BLWN gust excitation for a CCF of 0.3 demonstrating the effectiveness of the proposed density shaping controller in reducing wing root bending moment and torsion RMS and power difference ($\Delta_P$) compared to the uncontrolled (open-loop) response.

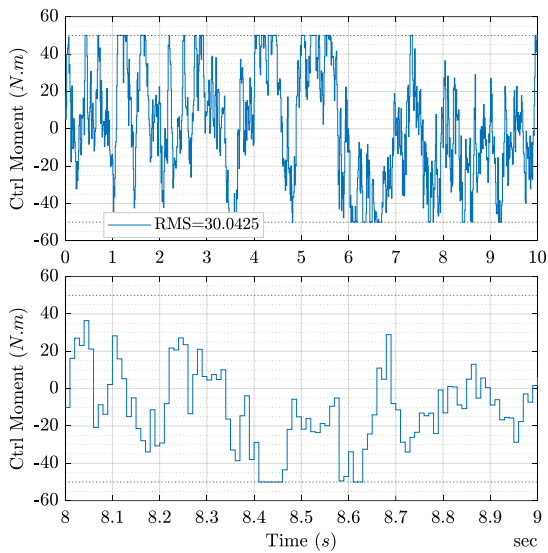
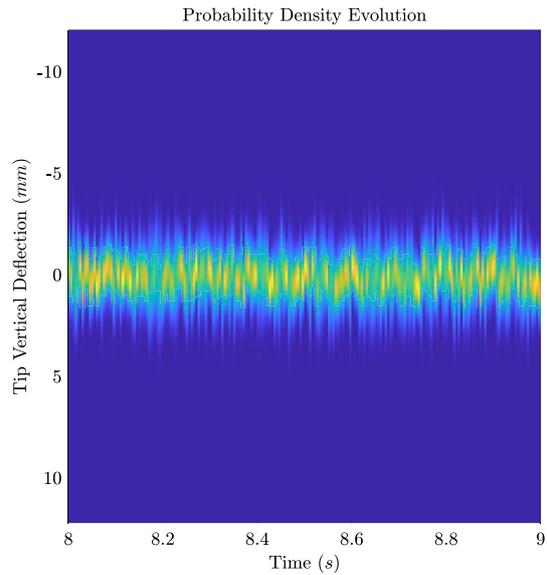

(a)        (b)

Figure 20 – (a) Control signal applied to the compliant chord modulates wing camber to suppress stochastic gust-induced vibrations, as illustrated by (b) the wing tip vertical deflection temporal evolution of probability density distribution under the Dryden gust profile with a CCF of 0.3 in response to the applied control

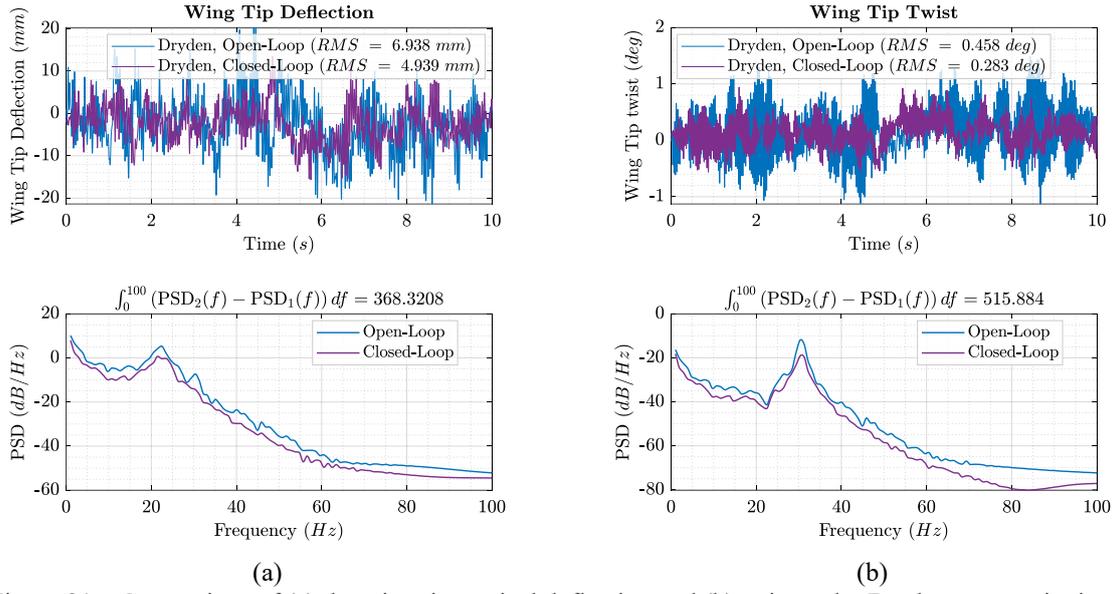

Figure 21 – Comparison of (a) the wing tip vertical deflection and (b) twist under Dryden gust excitation for a CCF of 0.3 demonstrating the effectiveness of the proposed density shaping controller in reducing wing tip deflection and twist RMS and power difference ($\Delta_P$) compared to the uncontrolled (open-loop) response.

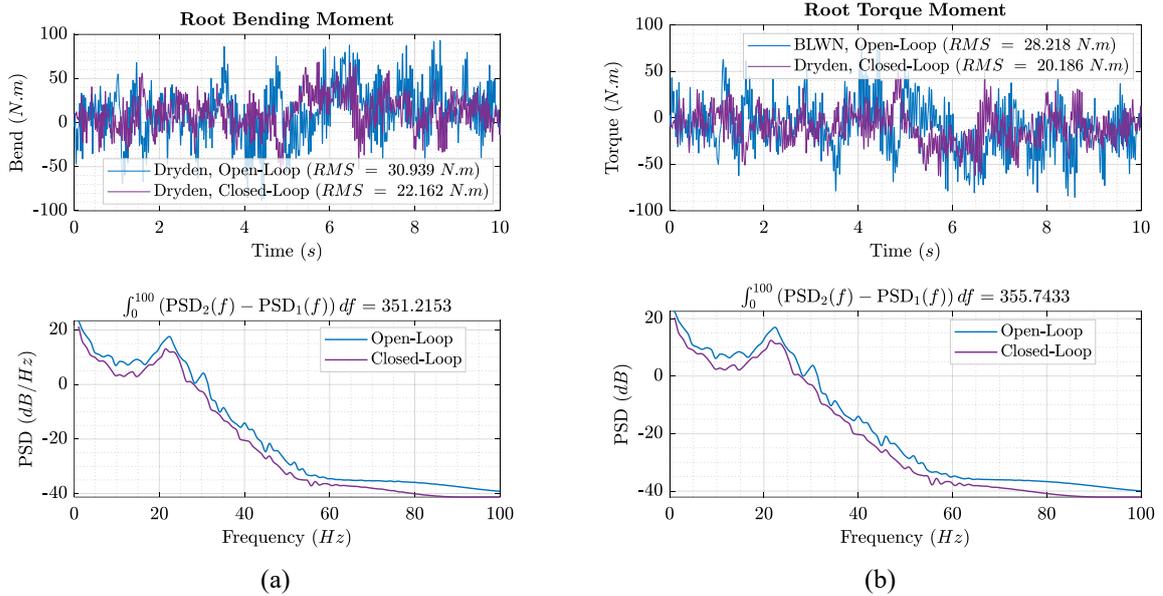

Figure 22 – Comparison of (a) the wing root bending and (b) torsional moments under Dryden gust excitation for a CCF of 0.3 demonstrating the effectiveness of the proposed density shaping controller in reducing wing root bending moment and torsion RMS and power difference ($\Delta_P$) compared to the uncontrolled (open-loop) response.